\documentclass[12pt]{article}

\makeatletter

\topmargin\z@

\def\section{\@startsection{section}{1}{\z@}{-\bigskipamount}{\bigskipamount}
{\bf}}

\newcommand{\bfm}[1]{\mathbf{#1}}

\newfont{\tts}{cmr5 scaled 1000}

\def\qed{\hfill$\square$}

\font\Bbb=msbm10 scaled\magstep1
\DeclareSymbolFont{AMSa}{U}{msa}{m}{n}
\DeclareSymbolFont{AMSb}{U}{msb}{m}{n}
\DeclareSymbolFontAlphabet{\mathbb}{AMSb}
\DeclareMathSymbol\square           {\mathord}{AMSa}{"03}

\def\Bbb{\mathbb}

\def\scs{\scriptstyle}
\def\scsc{\scriptscriptstyle}
\def\ds{\displaystyle}

\def\C{{\Bbb{C}}}
\def\N{{\Bbb{N}}}
\def\R{{\Bbb{R}}}

\begin{document}

\input prepictex
\input pictex
\input postpictex

\renewcommand\footnoterule{\noindent\rule{1.6cm}{0.025cm}\medskip}

\thispagestyle{empty}

\begin{center}
{\Large\bf Discrete phase space - I: \\
Variational formalism for classical \\[0.2cm]
relativistic wave fields} \\[1.2cm]
{\large\bf A. Das\footnote[1]{{\it E-mail address:} das@sfu.ca}}
\end{center}
\vskip0.6cm

\centerline{\it Department of Mathematics and Statistics}
\centerline{\it Simon Fraser University, Burnaby, B.C.,
V5A 1S6, Canada}

\vskip0.6cm

\noindent
\hrulefill
\vskip0.2cm

\renewcommand{\baselinestretch}{0.9}
\small\normalsize

\noindent
{\small\bf Abstract}\bigskip

\noindent
{\small
The classical relativistic wave equations are presented as partial
{\em difference\/} equations in the arena of covariant discrete phase
space. These equations are also expressed as {\em
difference-differential\/} equations in discrete phase space and
continuous time. The relativistic invariance and covariance of
the equations in {\em both\/} versions are established. The partial
difference and difference-differential equations are derived as the
Euler-Lagrange equations from the variational principle. The difference
and difference-differential conservation equations are derived. Finally,
the total momentum, energy, and charge of the relativistic classical
fields satisfying {\em difference-differential\/} equations are
computed.}
\vskip0.6cm


\noindent
\hrulefill
\vskip1cm

\renewcommand{\baselinestretch}{1}
\small\normalsize

\section{Introduction}

Partial difference equations have been studied [1] for a long
time to investigate problems in mathematical physics.
Moreover, modern numerical analysis [2], which studies the
differential equations arising out of various physical problems
approximately, is based upon finite difference (ordinary or
partial) equations.

Recently [3], we have formulated the wave mechanics in an
exact fashion in the arena of discrete phase space in terms of
the partial difference equations. This new formulation includes
(free) classical relativistic Klein-Gordon, Dirac, and gauge
field equations.

The proofs of the relativistic invariance or covariance of various
partial difference and difference-differential equations are {\em
quite subtle}. However, we have managed to provide such proofs in
Section~4.

The Euler-Lagrange equations for the partial difference and
difference-differ\-ential equations and Noether's theorems for
difference and difference-differential conservation laws are derived
in Appendix I and Appendix II respectively.

{F}inally, the total momentum, energy and charge are computed for
various relativistic classical fields obeying {\em
difference-differential\/} equations in Section~5. (We have
not computed conserved quantities for wave fields satisfying partial
difference equations for a physical reason to be explained later.)

Moreover, the stress-energy-momentum tensor and the consequent total
mo\-mentum-energy as computed from a general Lagrangian are somewhat
incomplete in this paper. However, in the next paper, these quantities
for the free Klein-Gordon, electro-magnetic, and Dirac fields are
furnished with {\em exact\/} equations.
\vskip1.2cm

\section{Notations and preliminary definitions}

There exists a characteristic length $\ell$ (which may be the
Planck length) in this theory. We choose the absolute units such that
$\hbar = c = \ell = 1.$ All physical quantities are expressed as
dimensionless numbers. Greek indices take from $\{1,2,3,4\},$ roman
indices take from $\{1,2,3\},$ and the capital roman take from
$\{1,2\}.$ Einstein's summation convention is followed in all three cases.
We denote the flat space-time metric by $\eta_{\mu \nu}$ and the diagonal
matrix $[\eta_{\mu \nu}] := {\mathrm{diag}} [1,1,1-1].$ (The signature
of the metric is obviously $+ 2.$) We denote the set of all
non-negative integers by $\N := \{0\} \cup \{Z^+\} =
\{0,1,2,3,\ldots\}.$ An element $n \in \N\times\N\times\N\times\N$
and an element $({\bfm{n}},t) \in \N^3 \times \R$ can be expressed
as
$$
n = (n^1, n^2, n^3, n^4), \;\; n^{\mu} \in \N, \;\; \mu \in
\{1,2,3,4\}\,;
\hspace*{0.75cm}
\eqno{\rm (1A)}
$$
$$
\hspace*{0.25cm}({\bfm{n}},t) = (n^1, n^2, n^3, t), \;\; n^j \in \N,
\;\; j \in \{1,2,3\}, \;\; t\in \R\vspace*{0.2cm}\,.
\eqno{\rm (1B)}
$$
Here and elsewhere, a bold roman letter indicates a {\em
three-dimensional\/} vector. In this paper, the equations in the
relativistic phase space are denoted by (..A), whereas the equations
in the discrete phase space and continuous time are labelled by (..B).
Both formulations are presented up to the difference and
difference-differential conservation laws. Subsequently, only the
difference-differential equations are pursued. The physical meanings
of the quantum numbers $n^\mu$ are understood from the equations
$$
(x^\mu )^2 + (p^\mu )^2 = 2n^\mu + 1, \quad \mu \in \{1,2,3,4\},
\eqno(2)
$$
where $x^\mu $ denote the space-time coordinates and $p^\mu $ stand
for four-momentum components in quantum mechanics.
\medskip

$$ \begin{minipage}[t]{15cm}
\beginpicture
\setcoordinatesystem units <0.65truecm,0.65truecm>
\setplotarea x from 0 to 13, y from 7 to -9

\setlinear
\plot 0 0 12 0 /
\plot 6 6 6 -6 /
\put {\vector(1,0){4}} [Bl] at 12 0
\put {\vector(0,1){4}} [Bl] at 6 6

\setquadratic
\circulararc 360 degrees from 5.35 0 center at 6 0
\circulararc 360 degrees from 4.75 0 center at 6 0
\circulararc 360 degrees from 4.15 0 center at 6 0

\setlinear
\plot 6 0  5.5 -0.3775 /
\plot 6 0  6.75 -1 /
\plot 6 0  7.075 1.5 /

\put {${\rm p}^\mu$} at 6 6.6
\put {${\rm q}^\mu$} at 12.7 0
\put {$\scriptscriptstyle 1$} at 5.8725 -0.35
\put {$\scriptscriptstyle \sqrt{3}$} at 6.75 -0.5255
\put {$\scriptscriptstyle \sqrt{5}$} at 7.05 1.075

\put {{\bf FIG. 1} \ \ One discrete phase plane.} at 6 -8.2
\endpicture
\end{minipage} $$
\vskip1cm

\noindent
Therefore, $n^\mu $ gives rise to a closed phase space loop of radius
$\sqrt{2n^\mu+1}$ in the $\mu$-th phase plane. Field quanta reside
in such phase space loops where the measurements of angles are {\em
completely uncertain\/} (see Fig.\,1). Phase space loops can be
interpreted as degenerate phase cells.

We shall encounter three dimensional improper integrals in computing
the total conserved quantities. Those integrals are always defined
to be the Cauchy-principal-value:
$$
\int\limits_{\R} f({\bfm{k}})\,d^3 {\bfm{k}} :=
\lim\limits_{L\rightarrow \infty} \left[\,\int\limits_{-L}^{L}
\int\limits_{-L}^{L} \int\limits_{-L}^{L} f(k_1, k_2,
k_3)\,dk_1dk_2dk_3\right]\vspace*{0.1cm}.
\eqno(3)
$$

Let a function be defined by $f: \N^4 \rightarrow \R$ (or $f: \N^4
\rightarrow \C$). Then the right partial difference, the left partial
difference, and the weighted-mean difference are defined respectively
by [3,4]\vspace*{0.1cm}:
$$
\Delta _\mu f(n) := f(.., n^\mu +1,..) - f(..,n^\mu ,..)
\vspace*{0.1cm}\,,
\eqno({\rm 4i})
$$
$$
\Delta^{\prime}_\mu f(n) := f(.., n^\mu ,..) - f(..,n^\mu -1,..)
\vspace*{0.2cm}\,,
\eqno({\rm 4ii})
$$
$$
\Delta_\mu^{\scsc\#} f(n) := (1 / \sqrt{2}\,)
\left[\sqrt{n^\mu +1}\,f (..,n^\mu +1,..) -\sqrt{n^\mu}\,f(..,n^\mu
-1,..)\right] \vspace*{0.3cm}
\eqno({\rm 4iii})
$$
It is clear that the partial difference operators $\Delta_\mu,
\Delta_\mu^{\prime}, \Delta_\mu^{\scsc\#}$ are all linear and the
operators $-i\Delta_\mu^{\scsc\#}$ are self-adjoint [3]. By direct
computations, we can prove the following generalizations of the
Leibnitz \vspace*{0.5cm}rule:
$$
\Delta _\mu [f(n) g(n)] = f(..,n^\mu +1,..)
\Delta_\mu g(n) + g(n) \Delta_\mu f(n)\vspace*{0.4cm}\,,
\eqno({\rm 5i})
$$
$$
\Delta_\mu^{\prime} [f(n) g(n)] = f(n) \Delta_\mu^{\prime} g(n) +
g(..,n^\mu-1,..) \Delta_\mu^{\prime} f(n)\vspace*{0.6cm}\,,
\eqno({\rm 5ii})
$$
$$
\begin{array}{rcl}
\Delta_\mu^{\scsc\#} [f(n) g(n)] &=& f(..,n^\mu\!+\!1,..)
\Delta_\mu^{\scsc\#} g(n) + g(..,n^\mu\!-\!1,..)
\Delta_\mu^{\scsc\#} f(n) \\[0.25cm]
&& - f(..,n^\mu\!+\!1,..) g(..,n^\mu\!-\!1,..) \Delta_\mu^{\scsc\#}
(1)\,, \\[0.35cm]
&=& f(..,n^\mu\!-\!1,..) \Delta_\mu^{\scsc\#} g(n)
+ g(..,n^\mu +1,..) \Delta_\mu^{\scsc\#} f(n) \\[0.25cm]
&& - f(..,n^\mu\!-\!1,..) g(..,n^\mu\!+\!1,..)
\Delta_\mu^{\scsc\#}(1)\vspace*{0.7cm}\,,
\end{array}
\eqno({\rm 5iii})
$$
$$
\Delta_\mu^{\scsc\#}(1) := (1 / \sqrt{2}\,) \left[
\sqrt{n^\mu +1} - \sqrt{n^\mu}\,\right], \; \lim\limits_{n^\mu
\rightarrow \infty} \Delta_\mu^{\scsc\#}(1) = 0\vspace*{0.5cm}\,,
\eqno({\rm 5iv})
$$
$$
\begin{array}{l}
\Delta_\mu \left\{\sqrt{n^\mu} \,[\phi (n) \psi(..,n^\mu-1,..)
+ \phi (..,n^\mu-1,..) \psi (n)]\right\} \\[0.3cm]
= \sqrt{2}\,[\phi (n) \Delta_\mu^{\scsc\#} \psi (n) + \psi (n)
\Delta_\mu^{\scsc\#} \phi (n)].
\end{array}
\eqno({\rm 5v})
$$
\vskip0.8cm

\noindent
In the left-hand side of the equation (5v), the index $\mu$ is {not}
summed.

We shall furnish a few more rules involving finite difference
operations in the following \vspace*{0.6cm}equations:
$$
\sqrt{n^\nu+1} \,\Delta_\nu \phi (n) + \sqrt{n^\nu}
\Delta_\nu^{\prime} \phi (n) = \sqrt{2}\,[\Delta_\nu^{\scsc\#}
\phi (n) - \phi (n)\cdot \Delta_\nu^{\scsc\#}(1)]\vspace*{0.7cm}\,,
\eqno({\rm 6i})
$$
$$
\begin{array}{l}
\sqrt{n^\nu+1} \,\Delta_\nu \Delta_\mu^{\scsc\#} \phi (n)
+ \sqrt{n^\nu} \Delta_\nu^{\prime} \Delta_\mu^{\scsc\#}
\phi (n) \\[0.3cm]
= \sqrt{2} \,[\Delta_\nu^{\scsc\#} \Delta_\mu^{\scsc\#}
\phi (n) - \Delta_\mu^{\scsc\#} \phi (n) \cdot
\Delta_\nu^{\scsc\#}(1)]\,,
\end{array}  \vspace*{0.8cm}\,
\eqno({\rm 6ii})
$$
$$
\begin{array}{l}
\sqrt{n^\mu+1} \,[\Delta_\mu \phi (n)]^2
- \sqrt{n^\mu} \,[\Delta_\mu^{\prime} \phi (n)]^2 \\[0.3cm]
= \sqrt{2} \{\Delta_\mu^{\scsc\#} [\phi (n)]^2 - 2\phi (n)
\Delta_\mu^{\scsc\#} \phi (n) + [\phi (n)]^2 \cdot
\Delta_\mu^{\scsc\#}(1)\}\,,
\end{array} \vspace*{0.4cm}\,
\eqno({\rm 6iii})
$$
$$
\begin{array}{l}
\sqrt{n^\mu+1} \,(\Delta_\mu \Delta_\nu^{\scsc\#}\phi)\cdot
(\Delta_\mu \Delta_\sigma^{\scsc\#}\phi) - \sqrt{n^\mu}
\,(\Delta_\mu^{\prime} \Delta_\nu^{\scsc\#}\phi)\cdot
(\Delta_\mu^{\prime} \Delta_\sigma^{\scsc\#} \phi) \\[0.3cm]
= \sqrt{2} \,\{\Delta_\mu^{\scsc\#}(\Delta_\nu^{\scsc\#}\phi\cdot
\Delta_\sigma^{\scsc\#}\phi) - \Delta_\nu^{\scsc\#}\phi\cdot
(\Delta_\mu^{\scsc\#}\Delta_\sigma^{\scsc\#}\phi)
- \Delta_\sigma^{\scsc\#}\phi\cdot (\Delta_\mu^{\scsc\#}
\Delta_\nu^{\scsc\#}\phi) \\[0.3cm]
+ (\Delta_\nu^{\scsc\#}\phi)\cdot (\Delta_\sigma^{\scsc\#}\phi)
\Delta_\mu^{\scsc\#}(1)\}\,,
\end{array} \vspace*{0.7cm}\,
\eqno{\rm (6iv)}
$$
$$
\begin{array}{l}
\sqrt{n^\mu+1} \,(\Delta_\mu A_\nu) (\Delta_\mu B_\sigma)
- \sqrt{n^\mu} \,(\Delta_\mu^{\prime}A_\nu) (\Delta_\mu^{\prime}
B_\sigma) \\[0.3cm]
= \sqrt{2}\,[\Delta_\mu^{\scsc\#}(A_\nu B_\sigma) - A_\nu
\Delta_\mu^{\scsc\#}B_\sigma - (\Delta_\mu^{\scsc\#}A_\nu) B_\sigma
+ A_\nu B_\sigma\Delta_\mu^{\scsc\#}(1)]\,.
\end{array}
\eqno({\rm 6v})
$$
\vskip0.6cm

\noindent
Here, neither the index $\mu$ nor the index $\nu$ is summed.

Now the rules for the summations will be listed\vspace*{0.4cm}.
$$
\sum\limits_{n^\mu=N_1^\mu}^{N_2^\mu} \Delta_\mu f(n)
= f(..,N_2^\mu+1,..) - f(..,N_1^\mu,..)\vspace*{0.4cm}\,,
\eqno({\rm 7i})
$$
$$
\sum\limits_{n^\mu=N_1^\mu}^{N_2^\mu} (\Delta_\mu^{\prime} f(n)
= f(..,N_2^\mu,..) - f(..,N_1^\mu-1,..)\vspace*{0.6cm}\,,
\eqno({\rm 7ii})
$$
$$
\begin{array}{l}
{\ds \sum\limits_{n^\mu=N_1^\mu}^{N_2^\mu} \Delta_\mu^{\scsc\#}f(n)
= (1 / \sqrt{2}) \Biggr[\sqrt{N_2^\mu+1} \:f(..,N_2^\mu +1,..)}
\\[0.2cm]
{\ds - \sqrt{N_1^\mu} \:f(..,N_1^\mu -1,..)
+ \sum\limits_{n^\mu = N_1^\mu+1}^{N_2^\mu} \sqrt{n^\mu}
\,\Delta_\mu^{\prime} f(n)\Biggr]\,.} \end{array}
\eqno({\rm 7iii})
$$
\vskip0.5cm

\noindent
It can be noted that right-hand sides of the equations (7i) and
(7ii) contain only boundary terms whereas the right-hand side of
(7iii) contains many more than just boundary terms.
\vskip1.4cm

\section{Gauss's theorem and conservation laws in a
discrete space}

Let a domain $D$ of the four-dimensional discrete space
$\N^4$ be given by (see Fig.\vspace*{0.6cm}\,2)
$$
D := \{n \in \N^4 : \,N_1^\mu < n^\mu < N_2^\mu , \; \mu \in
\{1,2,3,4\}\}\,.
\eqno(8)
$$

\vskip0.3cm

$$ \hspace*{-1cm}\begin{minipage}[t]{15cm}
\beginpicture
\setcoordinatesystem units <.65truecm,.65truecm>
\setplotarea x from 0 to 15, y from 7 to -7

\setsolid
\setlinear
\plot 0 0 14 0 /
\plot 6 6 6 -6 /
\put {\vector(1,0){4}} [Bl] at 14 0
\put {\vector(0,1){4}} [Bl] at 6 6

\setquadratic
\plot 7.85 1.1  7.7 0.8  8 0.5 /
\plot 8 0.5  8.25 0.2  8.3 -0.4 /
\plot 13.15 5.35 13.4 5.5 13.7 5.575 /

\setlinear
\put {\vector(1,0){4}} [Bl] at 13.7 5.575
\put {\vector(0,-1){4}} [Bl] at 8.305 -0.4

\put {${\rm n}^2$} at 6 6.7
\put {${\rm n}^1$} at 14.7 0.1
\put {$({\rm N}_1^1, {\rm N}_1^2)$} at 9.4 -0.9
\put {$({\rm N}_2^1, {\rm N}_2^2)$} at 15.1 5.6

\setplotsymbol ({\circle*{2.5}} [Bl])
\plotsymbolspacing8mm
\setdots <5mm> \plot 8 5.2  13.5 5.2 /
\plot 8 4.2  13.5 4.2 /
\plot 8 3.2  13.5 3.2 /
\plot 8 2.2  13.5 2.2 /
\plot 8 1.2  13.5 1.2 /
\setsolid

\put {{\bf FIG. 2} \ \ A two-dimensional dicrete domain.} at 7 -8.4
\endpicture
\end{minipage} $$
\vskip1.6cm

\noindent
The discrete boundary points of $D$ are taken to \vspace*{0.3cm}be
$$
\begin{array}{l}
\!\!\!\!\partial D = \partial_{1-}D \cup \partial_{1+}D \cup
\partial_{2-}D \cup \partial_{2+}D \cup \partial_{3-}D \cup
\partial_{3+}D \cup \partial_{4-}D \cup \partial_{4+}D , \\[0.4cm]
\!\!\!\!\partial_{\mu-}D := \{n\in \N^4 : \,n^\mu = N_1^\mu,
N_1^\sigma \leq n^\sigma \leq N_2^\sigma ,\sigma \neq \mu\} ,\\[0.4cm]
\!\!\!\!\partial_{\mu+}D := \{n\in \N^4 : \,n^\mu = N_2^\mu,
N_1^\sigma \leq n^\sigma \leq N_2^\sigma ,\sigma \neq \mu\} .
\end{array}
\eqno(9)
$$
\vskip0.7cm

We also denote the unit ``normal'' $\nu_\mu$ on the boundary $\partial
D$ by the following definition\vspace*{0.2cm}.
$$
\nu_\mu (n) := \left\{ \begin{array}{l}
\phantom{-} 1 \; \mbox{on} \; \partial_{\mu+} D\,, \\[0.2cm]
-1 \; \mbox{on} \; \partial_{\mu-} D\,.
\end{array} \right.
\eqno(10)
$$
\vskip0.8cm

We assume that a tensor field $j_{..}^{\mu ..}(n)$ is defined over
$D\subset \N^4.$ (See equation (34A) for the definition of a tensor
field.) Now we are ready to state and prove formally the ``discrete
Gauss's theorem'' [5].
\vskip1.4cm

\noindent
{\bf Theorem 3.1}
\,{\rm (Discrete Gauss's):} \quad {\it Let a tensor field
$j^{\mu ..}(n)$ be defined on $D\cup \partial D \subset \N^4.$ Then
$$
\begin{array}{rcl}
{\ds \sum\limits_{n^1=N_1^1}^{N_2^1} \sum\limits_{n^2=N_1^2}^{N_2^2}
\sum\limits_{n^3=N_1^3}^{N_2^3} \sum\limits_{n^4=N_1^4}^{N_2^4}
\,[\Delta_\mu j_{..}^{\mu ..}(n)] } &=:& {\ds \sum\limits_{D\subset
\N^4}^{(4)} \,[\Delta_\mu j_{..}^{\mu ..}(n)] } \\[0.6cm]
&=& {\ds \sum\limits_{\partial D\subset \N^4}^{(3)} j_{..}^{\mu ..}
(n) \nu_\mu (n)\,.}
\end{array}
\eqno(11)
$$
}
\vskip0.3cm

\noindent
{\bf Proof.} \quad
Using the equation (7i) four times to the left-hand side of the
equation (11) we \vspace*{0.2cm}obtain
$$
\begin{array}{l}
{\ds \sum\limits_{n^2=N_1^2}^{N_2^2} \sum\limits_{n^3=N_1^3}^{N_2^3}
\sum\limits_{n^4=N_1^4}^{N_2^4} \left[\,\sum\limits_{n^1=N_1^1}^{N_2^1}
\Delta_1 j_{..}^{1..} (n)\right] + .. + .. + .. } \\[0.7cm]
{\ds = \sum\limits_{n^2=N_1^2}^{N_2^2} \sum\limits_{n^3=N_1^3}^{N_2^3}
\sum\limits_{n^4=N_1^4}^{N_2^4} [\,j_{..}^{1..}(N_2^1,..)
- j_{..}^{1..} (N_1^1,..)] + .. + .. + .. } \\[0.7cm]
{\ds \phantom{=\hspace*{0.04cm}} \left[\,\sum\limits_{\partial_{1+}
D}^{(3)} j_{..}^{1..}(n) - \sum\limits_{\partial_{1-}D}^{(3)}
j_{..}^{1..} (n)\right] + .. + .. + .. } \\[0.7cm]
{\ds = \left[\,\sum\limits_{\partial_{1+}D\cup \partial_{1-}D}^{(3)}
j_{..}^{1..}(n) \nu_1(n)\right] + .. + .. + .. } \\[0.7cm]
{\ds = \sum\limits_{\partial D\subset \N^4}^{(3)} j_{..}^{\mu ..}
(n) \nu_\mu (n)\,. }
\end{array}
\eqno{\raisebox{-18.25ex}{\qed}}
$$
\vskip0.4cm

We shall now make some comments on the preceding theorem.\\

\noindent
(i) \ An alternate theorem holds by replacing the \,left-hand
side\, of the equa\-tion (11) by $\sum\limits_{D\subset \N^4}^{(4)}
{[\Delta_\mu^{\prime} j_{..}^{\mu ..}(n)]}.$
\vskip0.4cm

\noindent
(ii) \ Both forms of Gauss's theorem can be generalized to any
finite-di\-men\-sion\-al discrete space.
\vskip0.3cm

\noindent
(iii) \ In the finite difference representation of quantum
mechanics [3], the four momentum operators are furnished by $P_\mu
= -i\Delta_{\mu}^{\scsc\#}.$ These are consequences of the {\em
relativistic\/} representations. However, Gauss's theorem~3.1,
which uses $\Delta_\mu$ operators, is {\em non-relativistic}.
(The relativistic Gauss's theorem involving $\sum\limits_{D\subset
\N^4}^{(4)} [\Delta_\mu^{\scsc\#} j_{..}^{\mu ..}(n)]$ is not yet
solved. See the comments at the end of this section.) The partial
difference and difference-differential conservation equations
(non-relativistic) are furnished \vspace*{0.1cm}by
$$
\Delta_\mu j_{..}^{\mu ..} (n) = 0\vspace*{0.1cm}\,,
\eqno({\rm 12A})
$$
$$
\Delta_b j_{..}^{b..} ({\bfm{n}},t) + \partial_t j_{..}^{4..}
({\bfm{n}},t) = 0\vspace*{0.3cm}\,.
\eqno({\rm 12B})
$$
Difference conservation equations lead to summation conservations.
We shall presently state and prove a theorem about this topic.
\vskip0.7cm

\noindent
{\bf Theorem 3.2}
\,{\rm (Conserved sums):} \quad {\it Let a tensor field $j_{..}^{\mu
..(n)}$ satisfy the partial difference conservation $\Delta_\mu
j_{..}^{\mu ..} (n)=0$ in a domain $D\subset \N^4$ given in the
equation {\rm (8)}. Let furthermore the boundary
\vspace*{0.2cm}conditions
$$
j_{..}^{b..} (n) \nu_b (n)_{|\partial_{a+} D\cup \partial_{a-}D}
= 0 \vspace*{0.3cm}\,
\eqno(13)
$$
hold. Then the \vspace*{0.2cm}sum
$$
\begin{array}{l}
{\ds \sum\limits_{n^1=N_1^1}^{N_2^1}\sum\limits_{n^2=N_1^2}^{N_2^2}
\sum\limits_{n^3=N_1^3}^{N_2^3} \,[j_{..}^{4..} (n^1, n^2, n^3,
N_2^4)] } \\[0.7cm]
{\ds = \sum\limits_{n^1=N_1^1}^{N_2^1}\sum\limits_{n^2=N_1^2}^{N_2^2}
\sum\limits_{n^3=N_1^3}^{N_2^3} \,[j_{..}^{4..}(n^1,n^2,n^3,N_1^4)]
} \\[0.7cm]
{\ds = \sum\limits_{n^1=N_1^1}^{N_2^1}\sum\limits_{n^2=N_1^2}^{N_2^2}
\sum\limits_{n^3=N_1^3}^{N_2^3} \,j_{..}^{4..} (n^1,n^2,n^3,n^4) }
\end{array} \vspace*{0.3cm}\,
\eqno(14)
$$
for all $n^4$ satisfying $N_1^4 \leq n^4 \leq N_2^4.$
}
\vskip0.7cm

\noindent
{\bf Proof.} \quad
Using Gauss's theorem 3.1 and the difference conservation equation
(12A), we conclude that
$$
\begin{array}{rcl}
0 &=& {\ds \sum\limits_{D\subset \N^4}^{(4)}\,[\Delta_\mu j_{..}^{\mu ..}
(n)] } \\[0.7cm]
&=& {\ds \sum\limits_{\partial_{1+}D\cup \partial_{1-}D}^{(3)}
j_{..}^{1..} (n) \nu_1(n) + .. + .. + \sum\limits_{\partial_{4+}D\cup
\partial_{4-}D}^{(3)} j_{..}^{4..} (n) \nu_4 (n). }
\end{array}   \vspace*{0.2cm}\,
$$
Assuming the boundary conditions (13), the above equation yields
$$
\begin{array}{rcl}
0 &=& {\ds \sum\limits_{\partial_{4+}D\cup \partial_{4-}D}^{(3)}
j_{..}^{4..} (n) \nu_4(n) } \\[0.6cm]
&=& {\ds \sum\limits_{\partial_{4+}D}^{(3)}
j_{..}^{4..} (n^1,n^2,n^3,N_2^4) - \sum\limits_{\partial_{4-}D}^{(3)}
j_{..}^{4..} (n^1,n^2,n^3,N_1^4). }
\end{array} \vspace*{0.1cm}\,
$$
Considering Gauss's theorem for a proper subset of $D,$ the equation
(14) follows.
\qed
\vskip0.5cm

In the case of the difference conservation equation (12)
being valid for the denumerably infinite domain $\N^4,$ we can
derive conserved sums under suitable boundary conditions. Such
boundary conditions (sufficient) \vspace*{0.1cm}are
$$
\begin{array}{rcl}
\partial_{a-}D &:=& \left\{n\in \N^4 : \,n^a=0, \; 0\leq
n^b < \infty, \; a \neq b\right\}\,, \\[0.2cm]
\partial_{a+}D &:=& \left\{n\in \N^4 : \,n^a=M, \; 0\leq n^b
< \infty, \; a \neq b\right\}\,, \\[0.2cm]
&& \lim\limits_{M\rightarrow \infty} \left[j_{..}^{b..}(n)
\,\nu_b(n)\right]_{|\partial_{a-}D\cup\partial_{a+}D}=0\,;
\end{array} \vspace*{0.4cm}\,
\eqno{\rm (15A)}
$$
$$
\begin{array}{rcl}
\partial_{a-}D &:=& \left\{({\bfm{n}},t)\in \N^3 \times R :
\,n^a=0, \; 0\leq n^b < \infty, \; a \neq b\right\}\,, \\[0.2cm]
\partial_{a+}D &:=& \left\{({\bfm{n}},t)\in \N^3 \times R :
\,n^a=M, \; 0\leq n^b < \infty, \; a \neq b\right\}\,, \\[0.2cm]
&& \lim\limits_{M\rightarrow \infty} \left[j_{..}^{b..}({\bfm{n}},t)
\,\nu_b ({\bfm{n}},t)\right]_{|\partial_{a-}D\cup\partial_{a+}D}=0\,.
\end{array}
\eqno({\rm 15B})
$$
\vskip0.4cm

\noindent
Under boundary conditions (15A), the equation (14) yields the
following totally conserved quantities (generalized charges!)
$$
\begin{array}{rll}
{\ds Q_{..}^{..} = \sum\limits_{n^1=0}^{\infty}
\sum\limits_{n^2=0}^{\infty} \sum\limits_{n^3=0}^{\infty}
j_{..}^{4..} (n^1,n^2,n^3,n^4) } &=:& {\ds \sum\limits_{{\bfm{n}}
\in \N^3}^{(3)} j_{..}^{4..} ({\bfm{n}},n^4) } \\[0.6cm]
&=& {\ds \sum\limits_{{\bfm{n}}\in \N^3}^{(3)} j_{..}^{4..}
({\bfm{n}},2)\,. }
\end{array}
\eqno({\rm 16A})
$$
\vskip0.1cm

In the case of the difference-differential conservation equations
(12B) and the boundary conditions (15B), we can derive the totally
conserved quantities
$$
Q_{..}^{..} = \sum\limits_{{\bfm{n}}\in \N^3}^{(3)} j_{..}^{4..}
({\bfm{n}},t) = \sum\limits_{{\bfm{n}}\in \N^3}^{(3)} j_{..}^{4..}
({\bfm{n}},0)\vspace*{0.1cm}\,.
\eqno({\rm 16B})
$$

One may wonder why are we considering the {\em non-relativistic\/}
Gauss's theorem and the consequent conserved sums at all! In
Section~5, we shall prove that {\em the theorems of this
section can be used tactfully to elicit relativistic conserved
sums.}
\vskip1cm

\section{Relativistic covariance of partial difference and
dif\-fer\-ence-dif\-fer\-ential wave equations}

The relativistic invariance or covariance of our partial
difference or difference-differential equations is quite delicate.
The criterion used in this paper has been developed through many
papers [6] published in the last three decades. Let us start
with a very simple example of Poincar{\'e} transformations and the
invariance of the usual wave equation under that transformation.
Consider the infinitesimal time-translation:
$$
\begin{array}{l}
\widehat{x}^a  = x^a, \, \widehat{x}^4 = x^4 + \varepsilon^4\,,
\\[0.1cm]
x^a  = \widehat{x}^a, \, x^4 = \widehat{x}^4 - \varepsilon^4\,.
\end{array}
\eqno(17)
$$
There exists in the old frame a {\em different\/} event $(x^{\scsc\#})$
which has the {\em same\/} coordinates $(x)$ in the new frame. The
coordinates of $(x^{\scsc\#})$  from (17) are given by
$$
x^{\scsc\# a} = x^a, \, x^{\scsc\# 4} =
x^4-\varepsilon^4, \, \widehat{x}^{\scsc\# 4} = x^4\vspace*{0.1cm}\,.
\eqno(18)
$$
Consider the transformation rule for a scalar field $\phi (x)$ given
by
$$
\begin{array}{rcl}
\widehat{\phi}(\widehat{x}) &=& \phi (x)\,, \\[0.1cm]
\widehat{\phi}(x) &=& \phi (x^{\scsc\#}) = \phi
(x^1,x^2,x^3,x^4-\varepsilon^4)\vspace*{0.2cm}\,.
\end{array}
\eqno(19)
$$

Let $\phi (x)$ be a Taylor-expandible (or analytic) function. In that
case we can express (19) by Lagrange's formula as
$$
\begin{array}{rrl}
\widehat{\phi}(x) &=& \left[\exp (-\varepsilon^4 \partial_4)\right]
\phi (x) \\[0.2cm]
&=& \left[\exp (-i\varepsilon^4 (-i) \partial_4)\right]
\phi (x) \\[0.2cm]
&=& \phi (x) - \varepsilon^4 \partial_4 \phi (x)
+ 0\left[(\varepsilon^4)^2\right]\,, \\[0.2cm]
\partial_\mu &:=& {\ds \frac{\partial}{\partial x^\mu }\,.}
\end{array}
\eqno(20)
$$
\vskip0.2cm

Moreover, let $\phi (x)$ satisfy the usual wave equation
$$
\eta^{\mu\nu} \partial_\mu \partial_\nu \phi (x) = 0\,.
\eqno(21)
$$
\vskip0.1cm

In the new frame,
$$
\begin{array}{rcl}
\eta^{\mu\nu} \partial_\mu \partial_\nu \widehat{\phi}(x) &=&
\eta^{\mu\nu} \partial_\mu \partial_\nu \{[\exp (-\varepsilon^4
\partial_4)] \phi (x)\} \\[0.2cm]
&=& [\exp (-\varepsilon^4 \partial_4)]\,[\eta^{\mu\nu} \partial_\mu
\partial_\nu \phi (x)] = 0\,.
\end{array} \vspace*{0.1cm}\,
$$
The above equation demonstrates in an {\em unusual\/} manner the
relativistic invariance of the wave equation under an infinitesimal
time translation. We shall follow similar proofs in the sequel.

Let us now re-examine the very concept of relativistic invariance
or covariance. The relativistic covariance does {\em not\/} necessarily
imply that the space and time coordinates must be treated on the same
footing. Nor do the equations which treat space and time variables on
the equal footing automatically imply the relativistic covariance. For
example, let us consider the partial difference Klein-Gordon equation
{[7]} in the lattice space-time as
$$
\eta^{\mu\nu} \Delta_\mu \Delta_\nu^{\prime} \phi (n) - m^2 \phi
(n) = 0\,.
\eqno(22)
$$
This equation treats discrete space and time variables on the same
footing. However, the equation (22) is certainly {\em not\/} invariant
under the continuous Poincar{\'e} group ${\mathcal{I}}O(3,1)!$ (Although
there exists the lattice Lorentz group [8], a subgroup of the Lorentz
group $O(3,1),$ which leaves the lattice space-time invariant.)

Consider another example, namely the Friedmann-Robertson-Walker
cosmological model of the universe. The corresponding metric and the
``orthonormal'' tetrad are given by [9]:
$$
\begin{array}{rcl}
ds^2 &=& [R(t)]^2 [1+K\delta_{ab} x^ax^b]^{-2} [\delta_{ij} dx^i dx^j]
- (dt)^2\,, \\[0.3cm]
e_{(j)}^\mu &=& [R(t)]^{-1} [1+K\delta_{ab} x^ax^b] \delta_{(j)}^\mu\,,
\\[0.3cm]
e_{(4)}^\mu &=& \delta_{(4)}^\mu\,.
\end{array}
\eqno(23)
$$
The above metric and the corresponding tetrad do not treat the space
and time variables on equal footing. (Although these are extracted
as exact solutions of Einstein's general covariant equations.)
However, if we consider a suitably parametrized motion curve given
by a time-like geodesic, the appropriate Lagrangian and the
four-momentum are given by:
$$
\begin{array}{rcl}
L(..) &=& (m / 2) \left\{[R(t)]^2 [1+K\delta_{ab} x^ax^b]^{-2}
{[\delta_{ij} \dot{x}^i \dot{x}^j]} - (\dot{t}^2)\right\}\,, \\[0.3cm]
p_{(j)} &:=& {\ds \frac{\partial L(..)}{\partial \dot{x}^\mu}
\,e_{(j)}^\mu = m[R(t)]\,[1+K\delta_{ab} x^ax^b]^{-1} \delta_{ji}
\dot{x}^i\,, } \\[0.5cm]
p_{(4)} &:=& {\ds \frac{\partial L(..)}{\partial \dot{x}^\mu}
\,e_{(4)}^\mu = -m\dot{t}\,, } \\[0.4cm]
m &>& 0\,.
\end{array}
\eqno(24)
$$
Recalling that $g_{\mu\nu} (x) \dot{x}^\mu \dot{x}^\nu \equiv -1$
along a time-like geodesic, we obtain from (24) that
$$
\eta^{(\mu)(\nu)} p_{(\mu)} p_{(\nu)} = -m^2\vspace*{0.2cm}\,.
\eqno(25)
$$
Therefore, the {\em special relativistic\/} equation (25) holds among
the tetrad components of the four-momentum locally. In fact, in every
reasonable curved universe, with {\em any\/} admissible coordinate
system, the special relativity holds {\em locally\/} whether or not
the Poincar{\'e} group ${\mathcal{I}}O(3,1)$ is {\em globally\/}
admitted as Killing motion.

In a flat space-time, the first quantization  of the equation (25)
leads to the relativistic operator equation:
$$
{[\eta^{\mu\nu} P_\mu P_\nu + m^2 \,{\mathrm{I}}\,]}
\vec{\phi} = \vec{0}\,.
\eqno(26)
$$
Here, $\vec{\phi}$ and $\vec{0}$ are vectors in the tensor-product
{[3]} of the Hilbert spaces. The $P_\mu$'s represent four
self-adjoint, unbounded linear operators and ``I'' stands for the
identity operator. The equation (26) physically represents the
quantum mechanics of a massive, spin-less, free particle. Let us
explore the relativistic invariance of the operator equation (26).
The finite and infinitesimal versions of a Poincar{\'e} transformation
in the classical level are given respectively by:
$$
\begin{array}{l}
\widehat{x}^\mu  = a^\mu + \ell_\nu^\mu x^\nu\,, \\[0.15cm]
\eta_{\mu\nu} \ell_\alpha^\mu \ell_\beta^\nu = \eta_{\alpha\beta}\;;
\\[0.35cm]
\widehat{x}^\mu = x^\mu + \varepsilon^\mu + \varepsilon_\nu^\mu x^\nu\,,
\\[0.125cm]
\varepsilon_{\mu\nu} := \eta_{\mu\sigma} \varepsilon_\nu^\sigma =
-\varepsilon_{\nu\mu} + 0(\varepsilon^2)\,.
\end{array} \vspace*{0.2cm}\,
\eqno(27)
$$

However, there exist in a deeper level, the quantum Poincar{\'e}
transformations on quantum mechanical operators. In fact, there are
{\em two\/} possible quantum Poincar{\'e} transformations of operators
which involve the {\em same\/} unitary mapping. The Heisenberg-type
of Poincar{\'e} transformations [10] under the infinitesimal
version of (27) are furnished by:
$$
U(\varepsilon) := \exp \left\{-i \varepsilon^\mu P_\mu + (i/4)
\varepsilon^{\mu\nu} (Q_\mu P_\nu - Q_\nu P_\mu + P_\nu Q_\mu
- P_\mu Q_\nu )\right\},
\eqno({\rm 28iH})
$$
$$
Q_\mu P_\nu - P_\nu Q_\mu = i\eta_{\mu\nu} {\mathrm{I}}\vspace*{0.2cm}\,,
\eqno({\rm 28iiH})
$$
$$
\widehat{P}_\mu = U^{\dagger} (\varepsilon ) P_\mu U(\varepsilon)
= P_\mu - \varepsilon_\mu^\rho P_\rho + 0(\varepsilon^2)\vspace*{0.2cm}\,,
\eqno({\rm 28iiiH})
$$
$$
\widehat{Q}^\mu = U^{\dagger} (\varepsilon) Q^\mu U(\varepsilon)
= Q^\mu + \varepsilon^\mu {\mathrm{I}} + \varepsilon_\rho^\mu Q^\rho
+ 0(\varepsilon^2)\vspace*{0.1cm}\,,
\eqno({\rm 28ivH})
$$
$$
\widehat{\!\!\vec{\phi}} = \vec{\phi}\vspace*{0.2cm}\,.
\eqno({\rm 28vH})
$$
Here, the dagger denotes the hermitian conjugation.

In the Schroedinger-type of covariance [10], the abstract operators
and the state vectors transform under the infinitesimal version of
(27) as:
$$
\widehat{P}_\mu = P_\mu , \, \widehat{Q}^\mu = Q^\mu\vspace*{0.05cm}\,,
\eqno({\rm 28iS})
$$
$$
\widehat{\!\!\vec{\phi}} = U(\varepsilon) \vec{\phi}\vspace*{0.2cm}\,,
\eqno({\rm 28iiS})
$$
$$
\begin{array}{rcl}
\delta_L \vec{\phi} &:=& \widehat{\!\!\vec{\phi}} - \vec{\phi} \; = \;
\Big\{\!-i\,[\varepsilon^\mu P_\mu - (1/4) \varepsilon^{\mu\nu}
\\[0.1cm]
&& \!\!(Q_\mu P_\nu
- Q_\nu P_\mu + P_\nu Q_\mu - P_\mu Q_\nu )] + 0(\varepsilon^2)\Big\}
\vec{\phi}\,.
\end{array} \vspace*{0.3cm}\,
\eqno({\rm 28iiiS})
$$
Here, $U(\varepsilon)$ is the same as in (28iH). The vector
$\delta_L \vec{\phi}$ is called the Lie-variation of the vector
$\vec{\phi}.$

Note that in the two types of transformations, the expectation values
of a polynomial operator $F(P,Q)$ is given by:
$$
\begin{array}{rcl}
\langle\,\,\widehat{\!\!\vec{\phi}}_b |F(\widehat{P}, \widehat{Q})|\,\,
\widehat{\!\!\vec{\phi}}_a\rangle_H &=& \langle \vec{\phi}_b
|F[U^{\dagger}PU, U^{\dagger}QU]| \vec{\phi}_a\rangle \\[0.25cm]
&=& \langle \vec{\phi}_b| U^{\dagger} F(P,Q)U|
\vec{\phi}_a\rangle = \langle U\vec{\phi}_b| F(P,Q)|
U\vec{\phi}_a\rangle \\[0.2cm]
&=& \langle\,\,\widehat{\!\!\vec{\phi}}_b |F(\widehat{P},
\widehat{Q})|\,\,\widehat{\!\!\vec{\phi}}_a\rangle_{ _S}\vspace*{0.1cm}\,.
\end{array}
$$
(Here, $\langle \vec{\phi}_b | \vec{\phi}_a\rangle$ denotes the
inner-product of the two vectors $\vec{\phi}_b$ and $\vec{\phi}_a$.)
Therefore, physical quantities transform exactly in the same manner
under Heisenberg-type and Schroedinger-type of quantum covariance
rules. We shall follow the Schroedinger-type of covariance in this
paper.

It is well known [11] that the operator $P^\mu P_\mu,$
which is one of the Casimir operators of the Poincar{\'e} group
${\mathcal{I}} O(3,1),$ {\em commute\/} with all the generators
$P_\mu$ and $(1/4) (Q_\mu P_\nu - Q_\nu P_\mu + P_\nu Q_\mu
- P_\mu Q_\nu )$ of the Poincar{\'e} group. Therefore, we
obtain from (28\,i-iv\,S),
$$
\begin{array}{l}
{[\widehat{P}^\sigma \widehat{P}_\sigma + m^2\,{\mathrm{I}}\,]}\,\,\,
\widehat{\!\!\vec{\phi}} = [P^\sigma P_\sigma +m^2\,{\mathrm{I}}\,]
\,[\vec{\phi}+\delta_L \vec{\phi}] \\[0.3cm]
= [P^\sigma P_\sigma + m^2\,{\mathrm{I}}\,]\,[\delta_L
\vec{\phi}] \\[0.3cm]
= -i\,[\varepsilon^\mu P_\mu\!-\!(1/4)\varepsilon^{\mu\nu}
(Q_\mu P_\nu\!-\!Q_\nu P_\mu\!+\!P_\nu Q_\mu\!-\!P_\mu Q_\nu)] \times
\\[0.2cm]
\phantom{ = } \;[P^\sigma P_\sigma
\!+\!m^2\,{\mathrm{I}}\,] \vec{\phi}\!+\!0(\varepsilon^2) \\[0.3cm]
= 0(\varepsilon^2)\,.
\end{array} \vspace*{0.2cm}\,
\eqno(29)
$$
Therefore, the operator equation (29) is an actual proof of the
relativistic invariance (up to the 2nd order terms) for the operator
Klein-Gordon equation (26).

We can generalize the Lie-variation (28ivH) for an arbitrary
relativistic tensor (or spinor) operator $\vec{\phi}^{..}$.
The appropriate definition is given by:
$$
\begin{array}{l}
\widehat{\,\vec{\phi}^{..}} - \vec{\phi}^{..} = \delta_L
\vec{\phi}^{..} := -i\,\Big\{\varepsilon^\mu P_\mu -(1/4)
\varepsilon^{\mu\nu} \\[0.25cm]
\Big[(Q_\mu P_\nu - Q_\mu P_\nu + P_\nu
Q_\mu - P_\mu Q_\nu ) + 2S_{\mu\nu {..}}^{..}\Big]\Big\}
\vec{\phi}^{..} + 0(\varepsilon^2)\,,
\end{array} \vspace*{0.1cm}\,
\eqno(30)
$$
where $S_{\mu\nu {..}}^{..} = -S_{\nu\mu {..}}^{..}$ denotes the
``spin operator''.

The values of the entries for $S_{\mu\nu {..}}^{..}$ can be calculated
exactly from the {\em usual\/} tensor and spinor transformation rules.
{F}or example, if we consider a vector field $\vec{\phi}^{..}$ with
components $\phi^\alpha,$ then $S_{\mu\nu\beta}^\alpha =
\eta_{\nu\beta} \delta_{\mu}^\alpha - \eta_{\mu\beta}
\delta_\nu^\alpha .$

Now we shall state a necessary criterion for the relativistic
invariance of an operator equation.

{\em ``Let an infinite-dimensional Hilbert space vector
$\vec{\phi}^{..}$ represent a particle with zero or non-zero spin.
Let the corresponding Lie-variation $\delta_L \vec{\phi}^{..}$ be
given by the equation {\rm (30)}. In the case of the operator
equation for $\vec{\phi}^{..}$ being relativistic invariant or
covariant, the Lie-variation $\delta_L \vec{\phi}^{..}$ must satisfy
the same operator equation as $\vec{\phi}^{..}$ does {\rm (}up to
the {\rm 2}nd order terms{\rm )}.''}

Let us apply the above criterion of covariance on different
first-quantized equations arising out of {\em different
representations\/} of quantum mechanics. Firstly, consider the usual
Schroedinger representation of quantum mechanics, namely, $P_\mu
= -i\partial_\mu$ and $Q^\mu = x^\mu.$ The operator equation (26)
of the mass-shell constraint goes over into
$$
-\square \phi (x) + m^2 \phi (x) := -[\eta^{\mu\nu} \partial_\mu 
\partial_\nu
\phi (x) - m^2 \phi (x)] = 0\,.
$$
This is the usual Klein-Gordon equation. The Lie-variation under
${\mathcal{I}}O(3,1)$ is given by
$$
\delta_L \phi (x) = -\left[\varepsilon^\mu \partial_\mu - (1/4)
\varepsilon^{\mu\nu} (x_\mu \partial_\nu - x_\nu \partial_\mu
+ \partial_\nu x_\mu - \partial_\mu x_\nu)\right] \phi (x)
+ 0(\varepsilon^2)\,.
$$
Therefore,
$$
\begin{array}{l}
-\square \delta_L \phi (x) + m^2 \delta_L \phi (x) \\[0.2cm]
= -[\varepsilon^\mu \partial_\mu\!-\!(1/4)\varepsilon^{\mu\nu}
(x_\mu \partial_\nu\!-\!x_\nu \partial_\mu\!+\!\partial_\nu
x_\mu\!-\!\partial_\mu x_\nu)]
\,[-\square \phi (x)\!+\!m^2 \phi (x)]\!+\!0(\varepsilon^2) \\[0.2cm]
= 0 (\varepsilon^2)\,.
\end{array} \vspace*{0.2cm}\,
$$
Therefore, the relativistic invariance (up to the second order term)
is assured.

In the second example, let us consider the finite-difference
representation [3] of the quantum mechanics by putting
$$
P_\mu = -i\Delta_\mu^{\scsc\#} , \quad Q_\mu =
(1 / \sqrt{2}\,) (\Delta_\mu \sqrt{n^\mu} - \sqrt{n^\mu}
\Delta_\mu^{\prime} + 2\sqrt{n^\mu}\, {\mathrm {I}})\,.
$$
Here, the index $\mu$ is not summed. The operator equation (26)
yields
$$
\begin{array}{l}
\eta^{\mu\nu} \Delta_\mu^{\scsc\#} \Delta_\nu^{\scsc\#} \phi (n)
- m^2 \phi (n) = 0\,, \\[0.2cm]
n := (n^1,n^2,n^3,n^4) \in \N^4\,.
\end{array} \vspace*{0.2cm}\,
\eqno{\rm (31A)}
$$
This is the finite difference version of the Klein-Gordon equation.
The corresponding Lie-variation is given by:
$$
\begin{array}{l}
\delta_L \phi (n) = -\Big\{\varepsilon^\mu \Delta_\mu^{\scsc\#} \phi
(n) - (1/4\sqrt{2}) \varepsilon^{\mu\nu} \Big[(\Delta_\mu \sqrt{n^\mu}
- \sqrt{n^\mu} \Delta^{\prime}_{\mu} + 2\sqrt{n^{\mu}})
\Delta_\nu^{\scsc\#} \\[0.2cm]
-(\Delta_\nu \sqrt{n^\mu} - \sqrt{n^\nu} \Delta_\nu^{\prime}
+ 2\sqrt{n^\nu}) \Delta_\mu^{\scsc\#} + \Delta_\nu^{\scsc\#}
(\Delta_\mu \sqrt{n^\mu} - \sqrt{n^\mu} \Delta_\mu^{\prime} +
2\sqrt{n^\mu}) \\[0.2cm]
-\Delta_\mu^{\scsc\#} (\Delta_\nu \sqrt{n^\nu} - \sqrt{n^\nu}
\Delta_\nu^{\prime} + 2\sqrt{n^\nu})\Big] \phi (n) \Big\} +
0(\varepsilon^2)\vspace*{0.2cm}\,.
\end{array}
$$
Here, the indices $\mu ,\nu$ are summed. It can be proved by direct
computations that
$$
\eta^{\mu\nu} \Delta_\mu^{\scsc\#} \Delta_\nu^{\scsc\#} [\delta_L
\phi (n)] - m^2 [\delta_L \phi (n)] = 0(\varepsilon^2)\,.
$$
Thus the partial difference equation (31A) is indeed invariant (up
to the second order terms) under the continuous Poincar{\'e} group!

Now, as the third example, let us consider a {\em mixed
difference-differential\/} representation of the quantum mechanics by
choosing
$$
\begin{array}{l}
P_b = -i\Delta_b^{\scsc\#}, \, P_4 = -i\partial_4 \equiv
i\partial_t\,, \\[0.2cm]
Q_b = (1 / \sqrt{2})\,(\Delta_b \sqrt{n^b} - \sqrt{n^b}
\Delta_b^{\prime} + 2\sqrt{n^b}\,{\mathrm{I}}\,), \, Q^4
= x^4\equiv t\,.
\end{array}
$$
Here, the index $b$ is not summed. The operator equation (26) yields
the {\em difference-differential\/} version of the Klein-Gordon
equation:
$$
\begin{array}{l}
\delta^{ab} \Delta_a^{\scsc\#} \Delta_b^{\scsc\#} \phi ({\bfm{n}},t)
- (\partial_t)^2  \phi ({\bfm{n}},t) - m^2 \phi ({\bfm{n}},t) = 0\,,
\\[0.2cm]
({\bfm{n}},t) \in \N^3 \times \R\,.
\end{array}
\eqno{\rm (31B)}
$$
Here $\delta^{ab}$ is the Kronecker delta. We have for the
Lie-variation
$$
\begin{array}{l}
\delta_L \phi ({\bfm{n}},t) = \\[0.25cm]
-\Big\{\varepsilon^b \Delta_b^{\scsc\#}
+ \varepsilon^4 \partial_t - (1/4\sqrt{2})
\varepsilon^{ab} \Big[(\Delta_a \sqrt{n^a} -
\sqrt{n^a} \Delta_a^{\prime} + 2\sqrt{n^a})
\Delta_b^{\scsc\#}
\\[0.25cm]
-(\Delta_b \sqrt{n^b} - \sqrt{n^b} \Delta_b^{\prime} + 2\sqrt{n^b})
\Delta_a^{\scsc\#} + \Delta_b^{\scsc\#}(\Delta^a \sqrt{n^a} -
\sqrt{n^a} \Delta_a^{\prime} + 2\sqrt{n^a}) \\[0.25cm]
- \Delta_a^{\scsc\#} (\Delta_b \sqrt{n^b} - \sqrt{n^b}
\Delta_b^{\prime} + 2\sqrt{n^b})\Big] \phi ({\bfm{n}},t) \\[0.25cm]
+ (1 / \sqrt{2}) \varepsilon^{a4} \Big[(\Delta_a \sqrt{n^a}
- \sqrt{n^a} \Delta_a^{\prime}) \partial_t - t\Delta_a^{\scsc\#}
\Big]\Big\} \phi ({\bfm{n}},t)\,.
\end{array} \vspace*{0.2cm}\,
\eqno{\rm (32B)}
$$
Here, the indices $a, b$ are summed. We can prove by a long
calculation that
$$
\delta^{ab} \Delta_a^{\scsc\#} \Delta_b^{\scsc\#} [\delta_L \phi
({\bfm{n}},t)] - (\partial_t)^2 [\delta_L \phi ({\bfm{n}},t)] - m^2
[\delta_L \phi ({\bfm{n}},t)] = 0(\varepsilon^2)\,.
$$
In other words, the difference-differential equation (31B) is indeed
invariant (up to the second order terms) under the ten-parameter
continuous group ${\mathcal{I}}O(3,1)$!

As the last example, consider the Schroedinger's
difference-differential equation [3] for a free particle
(with $m>0)$ as
$$
(2m)^{-1} \delta^{ab} \Delta_a^{\scsc\#} \Delta_b^{\scsc\#} \psi
({\bfm{n}},t) + i\partial_t \psi ({\bfm{n}},t) = 0\,.
$$
The Lie-variation $\delta_L \psi({\bfm{n}},t)$ according to the
equation (32B) does {\em not\/} satisfy the Schroedinger
difference-differential equation up to the second order terms.
The reason for this failure is the operator $P_4 = i\partial_t$
does {\em not\/} commute with three particular generators $Q_4P_b
- Q_bP_4$ of the Poincar{\'e} group. Therefore, we obtain an
alternate proof of the well-known fact that the Schroedinger
equation is non-relativistic. Thus, our necessary criterion
involving the Lie-variation can prove or else disprove the
relativistic invariance or covariance (up to the second order
terms) of a quantum mechanical system in {\em any\/} representation.

We need to define the exact transformation rules for a tensor or
spinor field in partial difference and difference-differential
representations. Recall from (27) that the finite Poincar{\'e}
transformation is given by
$$
\begin{array}{l}
\widehat{\eta}^\mu = a^\mu + \ell_\nu^\mu x^\mu \,, \\[0.1cm]
\omega^{\mu\sigma} :=  \eta^{\sigma\nu} (\ell_\nu^\mu -
\delta_\nu^\mu)\,, \\[0.1cm]
\omega _{\alpha\beta} + \omega_{\beta\alpha} + \eta_{\mu\nu}
\omega_\alpha^\mu \omega_\beta^\nu = 0\,.
\end{array} \vspace*{0.2cm}\,
\eqno(33)
$$

The {\em exact\/} transformation rules for the tensor and spinor fields
under (33) are furnished by:
$$
\begin{array}{rcl}
\widehat{\phi}^{..}(n) &=& \Big(\exp \Big\{-a^\mu \Delta_\mu^{\scsc\#}+
(1/8\sqrt{2}) \,(\omega^{\mu\nu}-\omega^{\nu\mu}) \\[0.35cm]
&& \Big[(\Delta_\mu\sqrt{n^\mu}-\sqrt{n^\mu}
\Delta_\mu^{\prime}+2\sqrt{n^\mu}) \Delta_\nu^{\scsc\#} \\[0.3cm]
&& - (\Delta_\nu \sqrt{n^\nu} - \sqrt{n^\nu} \Delta_\nu^{\prime} +
2\sqrt{n^\nu}) \Delta_\mu^{\scsc\#} \\[0.3cm]
&& + \Delta_\nu^{\scsc\#} (\Delta_\mu
\sqrt{n^\mu} - \sqrt{n^\mu} \Delta_\mu^{\prime} + 2\sqrt{n^\mu})
\\[0.3cm]
&& - \Delta_\mu^{\scsc\#} (\Delta_\nu \sqrt{n^\nu} - \sqrt{n^\nu}
\Delta_\nu^{\prime} + 2\sqrt{n^\nu}) + i2S_{\mu\nu ..}^{..}
\Big]\Big\}\Big) \cdot \phi^{..}(n)\,,
\end{array}
\eqno{\rm (34A)}
$$
\vskip0.75ex
$$
\begin{array}{rcl}
\widehat{\phi}^{..}({\bfm{n}},t) &=& \Big(\exp \Big\{-a^b \Delta_b^{\scsc\#}
- a^4\partial_t - (1/8\sqrt{2})\,(\omega^{ab} - \omega^{ba}) \\[0.3cm]
&& \Big[(\Delta_a
\sqrt{n^a} - \sqrt{n^a} \Delta_a^{\prime} + 2\sqrt{n^a})
\Delta_b^{\scsc\#} \\[0.3cm]
&& - (\Delta_b \sqrt{n^b} - \sqrt{n^b} \Delta_b^{\prime} +
2\sqrt{n^b}) \Delta_a^{\scsc\#} \\[0.3cm]
&& + \Delta_b^{\scsc\#} (\Delta_a
\sqrt{n^a} - \sqrt{n^a} \Delta_a^{\prime} + 2\sqrt{n^a}) \\[0.3cm]
&& - \Delta_a^{\scsc\#} (\Delta_b \sqrt{n^b} - \sqrt{n^b}
\Delta_b^{\prime} + 2\sqrt{n^b}) + i2S_{ab ..}^{..}\Big] \\[0.3cm]
&& - (i/2\sqrt{2})\,
(\omega^{a4} - \omega^{4a})\Big[(\Delta_a \sqrt{n^a} - \sqrt{n^a}
\Delta_a^{\prime}) \partial_t \\[0.3cm]
&& - t\Delta_a^{\scsc\#} + iS_{a4 ..}^{..}\Big]\Big\}\Big) \cdot
\phi^{..}({\bfm{n}},t)\,.
\end{array}  \vspace*{0.3cm}\,
\eqno{\rm (34B)}
$$
Here, indices $\mu, \nu$ and $a, b$ are summed and $S_{\mu\nu ..}^{..}$
stands for the spin-operator for the field $\phi^{..}(n)$ or
$\phi^{..}({\bfm{n}},t).$

An obvious question that arises is how the discrete variables $n^1,
n^2,n^3,n^4$ transform from one inertial observer to another! We can
recall that the integer $n^\mu$ appears as the eigenvalue $2n^\mu+1$
of the operator $(P^\mu)^2 + (Q^\mu)^2.$ A particle in the
corresponding eigenstate $\vec{\psi}_{n^\mu }$ is located on a circle
of radius $\sqrt{2n^\mu+1}$ in the $p^\mu\!\!-\!x^\mu$-th phase plane
(see Fig.\,1).
However, for a relatively moving observer, according to the
Schroedinger-type of covariance, the particle is in the state
$\,\,\widehat{\!\!\vec{\psi}} = U(\varepsilon) \vec{\psi}_{n^\mu}.$
(See equations (28iS,iiS,iiiS).) \vspace*{0.075cm}\,\,But
$\,\,\widehat{\!\!\vec{\psi}}$
is {\em not\/} an eigenstate of the operator $(\widehat{P}^\mu)^2 +
(\widehat{Q}^\mu)^2 = (P^\mu)^2 + (Q^\mu)^2.$ Therefore, the particle
appears in a {\em fuzzy domain\/} in the phase plane for the moving
observer. But for the moving observer, there exists {\em another
state\/} $\,\,\widehat{\!\!\vec{\psi}}_{n^\mu} := \vec{\psi}_{n^\mu}
\neq U(\varepsilon) \vec{\psi}_{n^\mu} = \,\,\widehat{\!\!\vec{\psi}}$
such that $[(\widehat{P}^\mu)^2 + (\widehat{Q}^\mu)^2]
\,\,\widehat{\!\!\vec{\psi}}_{n^\mu} = [(P^\mu)^2 + (Q^\mu)^2]
\vec{\psi}_{n^\mu} = (2n^{\mu}+1) \,\,\widehat{\!\!\vec{\psi}}_{n^{\mu}}.$
Therefore, $\,\,\widehat{\!\!\vec{\psi}}_{n^\mu}$
is an eigenvector for the operator $(\widehat{P}^\mu)^2 +
(\widehat{Q}^\mu)^2$ in the moving frame and the corresponding
eigenvalue is $2n^\mu+1.$ {\em Thus both observers have exactly
similar discretized phase planes {\rm(}see Fig.\,{\rm 1),} although
discrete circles in one frame do not transform into discrete circles
in another frame!}

Now, we shall prove the relativistic invariance of the summation
operation $\sum\limits_{n^1=0}^{\infty} \sum\limits_{n^2=0}^{\infty}
\sum\limits_{n^3=0}^{\infty} \sum\limits_{n^4=0}^{\infty}$ and the
sum-integral operation $\sum\limits_{n^1=0}^{\infty}
\sum\limits_{n^2=0}^{\infty} \sum\limits_{n^3=0}^{\infty}\,
\int\limits_{\R} dt.$ Consider the state vector $\vec{\phi}$
representing a scalar particle in the abstract equations in (26) and
(28i,ii,iiiS). Mathematically speaking, $\vec{\phi} \in {\cal{H}}_1
\otimes {\cal{H}}_2 \otimes {\cal{H}}_3 \otimes {\cal{H}}_4$, is the
tensor product of Hilbert spaces [3,12]. Here, the Hilbert space
${\cal{H}}_\mu$ is acted upon by the operators $P_\mu, Q_\mu$ etc.
The square of the norm of $\vec{\phi}$ is given by:
$$
\|\,\vec{\phi}\,\|^2 := \langle
\vec{\phi}\,|\vec{\phi}\,\rangle\vspace*{0.2cm}\,.
$$
But from (28iH), (28iiS) we have
$$
\|\,\,\,\widehat{\!\!\vec{\phi}}\,\|^2 = \langle\vec{\phi}\,|U^{\dagger}
(\varepsilon) U(\varepsilon) \vec{\phi}\,\rangle
= \|\,\vec{\phi}\,\|^2\,,
$$
where the dagger denotes the hermitian-conjugation. Thus, the norm
$\|\,\vec{\phi}\,\|$ is invariant under the unitary transformation
induced by an infinitesimal Poincar{\'e} transformation. \\
Moreover, under a finite Poincar{\'e} transformation which induces
a unitary mapping, we can conclude that
$\|\,\,\,\widehat{\!\!\vec{\phi}}\,\| = \|\,\vec{\phi}\,\|.$

In the finite difference representation of the scalar field [3]
we have
$$
\|\,\vec{\phi}\,\|^2 := \sum\limits_{n^1=0}^{\infty}
\sum\limits_{n^2=0}^{\infty} \sum\limits_{n^3=0}^{\infty}
\sum\limits_{n^4=0}^{\infty} |\phi (n^1,n^2,n^3,n^4)|^2\vspace*{0.1cm}\,.
$$
Moreover, in the difference-differential representation
$$
\|\,\vec{\phi}\,\|^2 := \sum\limits_{n^1=0}^{\infty}
\sum\limits_{n^2=0}^{\infty} \sum\limits_{n^3=0}^{\infty}\,
\int\limits_{\R}\, |\phi (n^1,n^2,n^3,t)|^2 dt\,.
$$
Therefore, by the invariance of $\|\vec{\phi}\|^2,$ \vspace*{-0.1cm}both
operations $\sum\limits_{n^1=0}^{\infty} \sum\limits_{n^2=0}^{\infty}
\sum\limits_{n^3=0}^{\infty} \sum\limits_{n^4=0}^{\infty}$ and
$\sum\limits_{n^1=0}^{\infty} \sum\limits_{n^2=0}^{\infty}
\sum\limits_{n^3=0}^{\infty}\,\int\limits_{\R} dt$ must be
relativistically invariant.
\vskip1.2cm

\section{Variational formalism and conservation equations}

Let us consider a real-valued tensor field $A^{\mu ..}(n)$
in the difference representation and $A^{\mu ..}({\bfm{n}},t)$ in
the difference-differential representations. The action sum or the
action sum-integral for such fields are defined respectively by:
$$
\begin{array}{l}
{\mathcal{A}}(A^{\nu ..}) := {\ds \sum\limits_{n^1=N_1^1}^{N_1^2}
\sum\limits_{n^2=N_1^2}^{N_2^2} \sum\limits_{n^3=N_1^3}^{N_2^3}
\sum\limits_{n^4=N_1^4}^{N_2^4} } \\[0.6cm]
L\left(n; y^{\nu ..}\,;
y_\mu^{\nu ..}\right)_{|y^{\nu ..}=A^{\nu ..}(n), \, y_\mu^{\nu ..}
= \Delta_\mu^{{\mbox{\tts\#}}} A^{\nu ..}(n)} \:,
\end{array}
\eqno{\rm(35A)}
$$
\vskip0.2cm
$$
\begin{array}{l}
{\mathcal{A}}(A^{\nu ..}) := {\ds \sum\limits_{n^1=N_1^1}^{N_1^2}
\sum\limits_{n^2=N_1^2}^{N_2^2} \sum\limits_{n^3=N_1^3}^{N_2^3}\,
\int\limits_{t_1}^{t_2} } \\[0.6cm]
L\Big({\bfm{n}},t; y^{\nu ..}\,;
y_b^{\nu ..}, y_4^{\nu ..}\Big)_{|y^{\nu ..}=A^{\nu ..}
({\bfm{n}},t), \, y_b^{\nu ..} = \Delta_b^{{\mbox{\tts\#}}}
A^{\nu ..}, \, y_4^{\nu..} = \partial_t A^{\nu ..}}\,.
\end{array} \vspace*{0.6cm}\,
\eqno{\rm(35B)}
$$

The Euler-Lagrange equations under {\em zero\/} boundary variations
for $A^{\nu ..}(n)$ or $A^{\nu ..}({\bfm{n}},t)$ are given by (see
Appendix~I):
$$
\frac{\partial L(..)}{\partial y^{\nu ..}\,_{\!|..}}\, -
\Delta_\mu^{\scsc\#} \left\{\left[ \frac{\partial L(..)}{\partial
y_\mu^{\nu ..}}\right]_{|..}\right\} = 0\vspace*{0.2cm}\,,
\eqno{\rm (36A)}
$$
$$
\frac{\partial L(..)}{\partial y^{\nu ..}\,_{\!|..}}\, -
\Delta_b^{\scsc\#} \left\{\left[ \frac{\partial L(..)}{\partial
y_b^{\nu ..}}\right]_{|..}\right\} - \partial_t \left\{\left[
\frac{\partial L(..)}{\partial y_4^{\nu ..}}\right]_{|..}\right\}
= 0\vspace*{0.5cm}\,.
\eqno{\rm (36B)}
$$

In the case of a complex-value tensor field $\phi^{\nu ..}(n)$ and
$\phi^{\nu ..}({\bfm{n}},t),$ the action sum and sum-integral are
defined respectively \vspace*{0.2cm}by:
$$
\begin{array}{l}
{\mathcal{A}} (\phi^{\nu ..}, \overline{\phi^{\nu ..}}\,) :=
{\ds \sum\limits_{n^1=N_1^1}^{N_2^1} \sum\limits_{n^2=N_1^2}^{N_2^2}
\sum\limits_{n^3=N_1^3}^{N_2^3} \sum\limits_{n^4=N_1^4}^{N_2^4} }
\\[0.6cm]
L\Big(n; \rho^{\nu ..}, \overline{\rho^{\nu ..}}, \rho_\mu^{\nu ..},
\overline{\rho_\mu^{\nu ..}}\Big)
{\hspace*{-0.2cm}\mbox{\raisebox{-2.225ex}{$
{\begin{array}{l} {\scsc |\rho^{\nu ..}
= \phi^{\nu ..}(n), \,\overline{\rho^{\nu ..}} = \overline{\phi^{\nu
..}(n)} } \\[-0.15cm] { \scsc |\rho_\mu^{\nu ..}=\Delta_\mu^{\mbox{\tts\#}}
\phi^{\nu..}, \,\overline{\rho_\mu^{\nu ..}} = \Delta_\mu^{\mbox{\tts\#}}
\overline{\phi^{\nu ..}(n)} } \end{array}}\,,  $}}}
\end{array}
\eqno{\rm (37A)}
$$
\vskip0.2cm
$$
\begin{array}{l}
{\mathcal{A}} (\phi^{\nu ..}, \overline{\phi^{\nu ..}}\,) :=
{\ds \sum\limits_{n^1=N_1^1}^{N_2^1} \sum\limits_{n^2=N_1^2}^{N_2^2}
\sum\limits_{n^3=N_1^3}^{N_2^3}\, \int\limits_{t_1}^{t_2} }
\\[0.6cm]
L\Big({\bfm{n}},t; \rho^{\nu ..}, \overline{\rho^{\nu ..}}\,;
\rho_b^{\nu..}, \overline{\rho_b^{\nu ..}}, \rho_4^{\nu ..},
\overline{\rho_4^{\nu..}}\Big){\hspace*{-0.2cm}\mbox{\raisebox{-3.5ex}
{${\begin{array}{l}
\noindent{\scsc |\rho^{\nu ..} = \phi^{\nu ..}({\bfm{n}},t),
\,\rho^{\nu ..} \noindent=  \overline{\phi^{\nu..}({\bfm{n}},t)} }
\\[-0.1cm] { \scsc |\rho_b^{\nu ..} = \Delta_b^{\mbox{\tts\#}}
\phi^{\nu..}, \,\overline{\rho_b^{\nu ..}} = \Delta_b^{\mbox{\tts\#}}
\overline{\phi^{\nu ..}} } \\[-0.1cm] { \scsc |\rho_4^{\nu ..}
= \partial_t \phi^{\nu ..}, \,\overline{\rho_4^{\nu ..}} = \partial_t
\overline{\phi^{\nu ..}} } \end{array} }
$}}}  dt\,. \end{array}
\eqno{\rm (37B)}
$$
\vskip0.7cm

\noindent
Here, the bar stands for the complex-conjugation. \\
The corresponding Euler-Lagrange equations are:
$$
\hspace*{-3.2cm}\frac{\partial L(..)}{\partial
\rho^{\nu ..} {\scs |..}}\, - \Delta_\mu^{\scsc\#}
\left\{\left[ \frac{\partial L(..)}{\partial
\rho_\mu^{\nu ..}}\right]_{|..}\right\} = 0\vspace*{0.25cm}\,,
\eqno{\rm (38A)}
$$
$$
\hspace*{-3.2cm}\frac{\partial L(..)}{\partial
\overline{\rho^{\nu ..}}{\scs |..}}\,
- \Delta_\mu^{\scsc\#} \left\{\left[ \frac{\partial L(..)}{\partial
\overline{\rho_\mu^{\nu ..}}}\right]_{|..}\right\} = 0\vspace*{0.4cm}\,,
\eqno(38\overline{\rm A})
$$
$$
\frac{\partial L(..)}{\partial \rho^{\nu ..}{\scs |..}}\, -
\Delta_b^{\scsc\#} \left\{\left[ \frac{\partial L(..)}{\partial
\rho_b^{\nu ..}}\right]_{|..}\right\} - \partial_t \left\{\left[
\frac{\partial L(..)}{\partial \rho_4^{\nu ..}}\right]_{|..}\right\}
= 0\vspace*{0.4cm}\,,
\eqno{\rm (38B)}
$$
$$
\hspace*{1.1cm}\frac{\partial L(..)}{\partial
\overline{\rho^{\nu ..}{\scs |..}}}\, -
\Delta_b^{\scsc\#} \left\{\left[ \frac{\partial L(..)}{\partial
\overline{\rho_b^{\nu ..}}}\right]_{|..}\right\} - \partial_t
\left\{\left[ \frac{\partial
L(..)}{\partial \overline{\rho_4^{\nu ..}}}\right]_{|..}\right\}
= 0\vspace*{0.5cm}\,.
\eqno(38\overline{B})
$$

In the derivation of above equations, the techniques of the
complex-conjugate coordinates are used [13,14].

We shall now derive the partial difference and the
difference-differential conservation equations for various fields
(see Appendix~II).
$$
\Delta_\mu T_\mu^\nu + .. = 0\vspace*{0.3cm}\,,
\eqno{\rm (39Ai)}
$$
$$
\begin{array}{rcl}
T_\mu^\nu (n) &:=& {\ds \sqrt{\frac{n^\nu}{2}} \:\Biggr[
\frac{\partial L(..)}{\partial y_\nu^{\alpha ..}}_{|(..,n^\nu-1,..)}
\cdot \Delta_\mu^{\scsc\#} A^{\alpha ..} } \\[0.5cm]
&& {\ds + \frac{\partial (L..)}{\partial
y_\nu^{\alpha ..}\,_{|..}} \cdot (\Delta_\mu^{\scsc\#}
A^{\alpha ..})_{|(..,n^\nu-1,..)} -
\delta_\nu^\mu  L(..)_{|..} \Biggr],}
\end{array} \vspace*{0.5cm}\,
\eqno{\rm (39Aii)}
$$
$$
\Delta_b T_a^b + \partial_t T_a^4 + .. = 0\vspace*{0.4cm}\,,
\eqno{\rm (39Bi)}
$$
$$
\Delta_b T_4^b + \partial_t T_4^4 = 0\vspace*{0.5cm}\,,
\eqno{\rm (39Bii)}
$$
$$
\begin{array}{rcl}
T_a^b ({\bfm{n}},t) &:=&  {\ds \sqrt{\frac{n^b}{2}} \:\Biggr[
\frac{\partial L(..)}{\partial y^{\alpha ..}}_{b|(..,n^b-1,..)}
\cdot \Delta_a^{\scsc\#} A^{\alpha}) } \\[0.5cm]
&& {\ds + \frac{\partial (L..)}{\partial
y_{\alpha..}}_{b|..} \cdot (\Delta_a^{\scsc\#}
A^{\alpha ..})_{|(..,n^b-1,..)} -
\delta_a^b  L(..)_{|..} \Biggr], }
\end{array} \vspace*{0.5cm}\,
\eqno{\rm (39Biii)}
$$
$$
T_a^4 ({\bfm{n}},t) := \frac{\partial L(..)}{\partial
y_{4|..}^{\alpha..}} \cdot \Delta_a^{\scsc\#}
A^{\alpha..}\vspace*{0.5cm}\,,
\eqno{\rm (39Biv)}
$$
$$
T_4^b ({\bfm{n}},t) := \sqrt{\frac{n^b}{2}} \:\Biggr[
\frac{\partial L(..)}{\partial y^{\alpha ..}_{b|(..,n^b-1,..)}}
\cdot \partial_t A^{\alpha ..}
+ \frac{\partial L(..)}{\partial
y_{b|..}^{\alpha ..}} \cdot (\partial_t A^{\alpha
..})_{|(..,n^b-1,..)} \Biggr],
\vspace*{0.5cm}\,
\eqno{\rm (39Bv)}
$$
$$
T_4^4 ({\bfm{n}},t) := \frac{\partial L(..)}{\partial y_{4|..}^{\alpha..}}
\cdot \partial_t A^{\alpha ..} - L(..)_{|..}\vspace*{0.5cm} \,.
\eqno{\rm (39Biv)}
$$
We can sum or sum-integrate the relativistic conservation equations
over an appropriate domain of $\N^4$ or $\N^3 \times \R.$ Using
Gauss's theorem~3.1 and {\em assuming\/} boundary conditions similar
to the equations (15A,B), we obtain the total conserved four-momentum:
$$
-P_\mu = \sum\limits_{{\bfm{n}}=0}^{\infty (3)} [T_\mu^4 ({\bfm{n}},n^4)
+ ..] = \sum\limits_{{\bfm{n}}=0}^{\infty (3)} [T_\mu^4 ({\bfm{n}},2)
+ ..]\vspace*{0.5cm}\,,
\eqno{\rm (40A)}
$$
$$
-P_b = \sum\limits_{{\bfm{n}}=0}^{\infty (3)} [T_b^4 ({\bfm{n}},t)
+ ..] = \sum\limits_{{\bfm{n}}=0}^{\infty (3)} [T_b^4 ({\bfm{n}},0)
+ ..]\vspace*{0.5cm}\,,
\eqno{\rm (40Bi)}
$$
$$
H := -P_4 = \sum\limits_{{\bfm{n}}=0}^{\infty (3)} [T_4^4 ({\bfm{n}},t)
+ ..] = \sum\limits_{{\bfm{n}}=0}^{\infty (3)} [T_4^4 ({\bfm{n}},0)
+ ..] \vspace*{0.5cm}\,,
\eqno{\rm (40Bii)}
$$

\noindent
$\sum\limits_{{\bfm{n}}=0}^{\infty (3)} := \sum\limits_{n^1=0}^{\infty}
\sum\limits_{n^2=0}^{\infty} \sum\limits_{n^3=0}^{\infty}.$
\vskip0.25cm

The total energy-momentum components $P_\mu$ are given by incomplete
equations (40A), (40Bi,ii) for a general Lagrangian. However, in
the following paper, we shall derive {\em exact\/} equations for the
Klein-Gordon, electro-magnetic, and Dirac fields.

In case of a complex-valued field $\phi^{\alpha..},$ the difference
and difference-differ\-ent\-ial conservation for the charge-current
vector $j^\mu$ is given by equations (A.II.9A,B) as:
$$
\Delta_{\mu} j^\mu (n) = 0\vspace*{0.4cm}\,,
\eqno{\rm (41Ai)}
$$
$$
\begin{array}{rcl}
j^\mu (n) &:=& {\ds ie \sqrt{\frac{n^\mu}{2}} \:\Biggr\{ \Biggr[
\frac{\partial L(..)}{\partial \rho_{\mu|(..,n^\mu-1,..)}^{\alpha ..}}
\cdot \phi^{\alpha ..}(n) } \\[0.5cm]
&& {\ds + \frac{\partial L(..)}{\partial
\rho_{\mu |..}^{\alpha ..}} \cdot \phi^{\alpha ..} (..,n^\mu-1,..)
\Biggr]\Biggr\} + {\rm (c.c.)} }\vspace*{0.6cm}\,,
\end{array}
\eqno{\rm (41ii)}
$$
$$
\Delta_b j^b ({\bfm{n}},t) + \partial_t j^4 ({\bfm{n}},t) =
0\vspace*{0.5cm}\,,
\eqno{\rm (41Bi)}
$$
$$
\begin{array}{rcl}
j^b ({\bfm{n}},t) &:=& {\ds ie \sqrt{\frac{n^b}{2}} \:\Biggr\{ \Biggr[
\frac{\partial L(..)}{\partial \rho^{\alpha ..}}_{b|(..,n^b-1,..)} \cdot
\phi^{\alpha ..}({\bfm{n}},t) } \\[0.5cm]
&& {\ds + \frac{\partial L(..)}{\partial
\rho_{b|..}^{\alpha ..}} \cdot \phi^{\alpha ..} (..,n^b-1,..)
\Biggr]\Biggr\} + {\rm (c.c.)}\,,}
\end{array}  \vspace*{0.5cm}\,
\eqno{\rm (41Bii)}
$$
$$
j^4 ({\bfm{n}},t) := ie \Biggr[\frac{\partial L(..)}{\partial
\rho_{4|..}^{\alpha ..}} \cdot \phi^{\alpha ..} ({\bfm{n}},t)
- \frac{\partial L(..)}{\partial\,\overline{\rho}_{4|..}^{\alpha ..}}
\cdot \overline{\phi^{\alpha ..}({\bfm{n}},t)}\Biggr]\vspace*{0.5cm}.
\eqno{\rm (41Biii)}
$$
Here, (c.c) stands for the complex-conjugation of the {\em preceding\/}
terms and $e = \sqrt{4\pi / 137}$ is the charge parameter.

Under appropriate boundary conditions (15A,B), we can derive the
conserved total charge:
$$
Q\!=\!-\frac{ie}{\sqrt{2}}
\sum\limits_{{\bfm{n}}=0}^{\infty (3)}
\left\{\!\left[\frac{\partial L(..)}{\partial
\rho_{4|({\bfm{n}},1)}^{\alpha ..}} \cdot \phi^{\alpha ..}
({\bfm{n}},2)\!+\!\frac{\partial L(..)}{\partial
\rho_{4|({\bfm{n}},2)}^{\alpha ..}} \cdot \phi^{\alpha ..}
({\bfm{n}},1) \right]\!\right\}\!+\!{\rm (c.c.)}\vspace*{0.6cm},\;
\eqno{\rm (42A)}
$$
$$
Q = \Biggr[-ie \sum\limits_{{\bfm{n}}=0}^{\infty (3)}
\frac{\partial L(..)}{\partial \rho_{4|({\bfm{n}},0)}^{\alpha ..}}
\cdot \phi^{\alpha ..} ({\bfm{n}},0)\Biggr]
+ {\rm (c.c.)}\vspace*{0.6cm}\,.
\eqno{\rm (42B)}
$$

Note that equations (42A,B) are {\em already exact}.
\vskip0.075cm

We shall now make some comments about the total conserved quantities.
In the preceding section, we have deduced that the four-fold summation
$\sum\limits_{n^1=0}^{\infty} \sum\limits_{n^2=0}^{\infty}
\sum\limits_{n^3=0}^{\infty} \sum\limits_{n^4=0}^{\infty}$
and the \vspace*{0.1cm}summation-integration $\sum\limits_{n^1=0}^{\infty}
\sum\limits_{n^2=0}^{\infty} \sum\limits_{n^3=0}^{\infty}
\,\int\limits_{\R} dt$ are {\em relativistically invariant\/} operations.
We have obtained the total four-momentum $P_\mu$ in the equations
(40A), (40Bi,ii)  by the four-fold summation and the
summation-integration of the {\em relativistic\/} conservation equations
(A.II.4A) and (A.II.4Bi,ii) respectively. Therefore, we claim that
(the complete) $P_\mu$'s are {\em components of a relativisitic
four-vector}. Similarly, the total charge in equations (42A) or
(42B) is {\em relativistic invariant}.

The physical contents of the partial difference conservation (39Ai)
and the difference-differential conservation (39Bi) are {\em
identical}. However, the total four-momentum components $P_\mu$ in
(40A) and (40Bi,ii) are {\em quite different\/} inspite of the same
notations! The reason for this distinction is that in equation (40A),
the summation of the field is over a $n^4 = {\mathrm{const.}}$
``hyperspace''. In covariant phase space, an $n^4={\mathrm{const.}}$
``hypersurface'' implies that $(t)^2 + (p_4)^2 = (2n^4+1) =
{\mathrm{const.}}$ Therefore, in the $t$-$p^4$ phase plane, the time
coordinate $t$ oscillates between the values $-\sqrt{2n^4+1}$ and
$\sqrt{2n^4+1}.$ It is certainly {\em not\/} a $t={\mathrm{const.}}$
``slice''. However, in the case of equations (40Bi,ii), the {\em
usual\/} $t = {\mathrm{const.}}$ ``hypersurface'' of the discrete phase
space and continuous time is used.
\vskip1.2cm

\section*{Appendix I: \, Euler-Lagrange equations}

We shall consider a {\em two-dimensional\/} lattice function
$f$ for the sake of simplicity. We firstly choose a two-dimensional
discrete domain (see Fig.\,2)
$$
D\!:=\!\{n\!\in\!\N^2\!\!: 0\!<\!N_1^A\!\!<\!N_2^A,
2\!\leq\!N_2^A\!\!-\!N_1^A,
N_1^A\!\!<\!n^A\!<\!N_2^A, A\!\in\!\{1,2\}\}.
\eqno{\rm (A.I.1)}
$$
We shall follow the summation convention on capital indices which
take values from $\{1,2\}.$

A real-valued lattice function $f: \,D\subset \N^2 \rightarrow \R$
is considered. The Lagrangian function $L: \,\tilde{D} \subset \N^2
\times \R \times \R^2 \rightarrow \R$ is assumed to be
Taylor-expandible (or real analytic). (In all practical purposes,
$L$ is a polynomial function.) The action functional ${\mathcal{A}}$
is defined to be the double-sum
$$
{\mathcal{A}} (f) := {\ds \sum\limits_{n^1=N_1^1}^{N_2^1}
\sum\limits_{n^2=N_1^2}^{N_2^2} L(n; y; y_A)_{|y=f(n),
\, y_A=\Delta_A^{\mbox{\tts\#}} f(n)} \;, }
\quad n := (n^1, n^2)\vspace*{0.3cm}\,.
\eqno{\rm (A.I.2)}
$$

Now we define the variations of the lattice function $f$ by
$$
\begin{array}{rcl}
\delta f(n) &:=& \varepsilon h(n)\,, \\[0.3cm]
\delta [\Delta_A^{\scsc\#} f(n)] &:=& \varepsilon \Delta_A^{\scsc\#}
h(n) = \Delta_A^{\scsc\#}[\delta f(n)]\,.
\end{array} \vspace*{0.2cm}\,
\eqno{\rm (A.I.3)}
$$
Here, $\varepsilon >0$ is an arbitrary, small, positive number.
Therefore, the variation of the action functional ${\mathcal{A}}$
is furnished by:
$$
\begin{array}{rcl}
\delta {\mathcal{A}} (f) &=& {\ds \sum\limits_{n^=N_1^1}^{N_2^1}
\sum\limits_{n^2=N_1^2}^{N_2^2} \Big\{ L(n;y;y_A)_{|y=f+\delta f,
\, y_A=\Delta_A^{\mbox{\tts\#}} f +\delta (\Delta_A^{\mbox{\tts\#}}
f)} } \\[0.6cm]
&& - L(n;y;y_A)_{|y=f(n), \, y_A=\Delta_A^{\mbox{\tts\#}}f(n)}\Big\}
\\[0.6cm]
&=& {\ds \varepsilon \sum\limits_{n^1=N_1^1}^{N_2^1}
\sum\limits_{n^2=N_1^2}^{N_2^2} \Biggr\{\Biggr[\frac{\partial
L(..)}{\partial y}\Biggr]_{|y=f(n),
\, y_A=\Delta _A^{\mbox{\tts\#}}f} } \\[0.7cm]
&& {\ds \cdot \,h(n) + \Biggr[ \frac{\partial L(..)}{\partial y_A}
\Biggr]_{|y=f(n), \, y_A=\Delta _A^{\mbox{\tts\#}}f} \cdot
\Delta_A^{\scsc\#} h(n) \Biggr\} + 0(\varepsilon^2)\,. }
\end{array} \vspace*{0.3cm}\,
\eqno{\rm (A.I.4)}
$$

The variational principle, which implies the stationary (or critical)
values of the action functional, can be stated succinctly as
$$
\lim\limits_{\varepsilon\rightarrow 0_+} \frac{\delta
{\mathcal{A}}(f)}{\varepsilon} = 0\,.
$$
The above equation yields from (A.I.4) that
$$
\sum\limits_{n^1=N_1^1}^{N_2^1} \sum\limits_{n^2=N_1^2}^{N_2^2}
\,\Biggr\{\Biggr[\frac{\partial L(..)}{\partial y}\Biggr]_{|(n^1,
n^2)} \!\!\cdot h(n)+\Biggr[\frac{\partial L(..)}{\partial
y_A}\Biggr]_{|(n^1, n^2)} \!\!\cdot \Delta_A^{\scsc\#} h(n)
\Biggr\} = 0\vspace*{0.2cm}.
\eqno{\rm (A.I.5)}
$$
(Here, we have simplified the notation by putting $|(n^1, n^2)$ in
place of $|y=f(n^1, n^2), \; y_A=\Delta_A^{\scsc\#} f(n^1, n^2)$.)
Using the definition (4iii), opening up the double-sum in (A.I.5),
and rearranging terms, we obtain (after a very long calculation)
$$
\!\begin{array}{l}
{\ds\sum\limits_{n^1=N_1^1+1}^{N_2^1-1}\,\sum\limits_{n^2=N_1^2
+1}^{N_2^2-1} \,\Biggr\{\Biggr[\frac{\partial L(..)}{\partial y}
\Biggr]_{|(n^1, n^2)} \!\!\!-\!\Delta_A^{\scsc\#}\Biggr[\frac{\partial
L(..)}{\partial y_A}\Biggr]_{|(n^1, n^2)}\Biggr\}
\cdot h(n^1, n^2) } \\[0.8cm]
+ \mbox{(Boundary terms)} \; = 0\,.
\end{array}\!\!   \vspace*{0.4cm}\,
\eqno{\rm (A.I.6)}
$$
Here, the boundary terms are given \vspace*{0.2cm}by:
$$
\begin{array}{l}
\mbox{(Boundary terms)} := \\[0.4cm]
{\ds -\sqrt{\frac{N_1^2}{2}} \,\sum\limits_{k=0}^{N_2^1-N_1^1}
\frac{\partial L(..)}{\partial y_2}_{|(N_1^1+k, N_1^2)} \cdot
h(N_1^1 + k, N_1^2-1) } \\[0.7cm]
{\ds +\sqrt{\frac{N_2^1+1}{2}} \,\sum\limits_{k=0}^{N_2^2-N_1^2}
\frac{\partial L(..)}{\partial y_1}_{|(N_2^1+1, N_1^2+k)} \cdot
h(N_2^1 + 1, N_1^2 + k) } \\[0.7cm]
{\ds +\sqrt{\frac{N_2^2+1}{2}} \,\sum\limits_{k=0}^{N_2^1-N_1^1}
\frac{\partial L(..)}{\partial y_2}_{|(N_1^1+k, N_2^2)} \cdot
h(N_1^1 + k, N_2^2 + 1) } \\[0.7cm]
{\ds -\sqrt{\frac{N_1^1}{2}} \,\sum\limits_{k=0}^{N_2^2-N_1^2}
\frac{\partial L(..)}{\partial y_1}_{|(N_1^1, N_1^2+k)} \cdot
h(N_1^1 - 1, N_1^2 + k) } \\[0.7cm]
{\ds +\sum\limits_{j=1}^{N_2^1-N_1^1-1}
\left\{\frac{\partial L(..)}{\partial y}_{|(N_1^1+j, N_1^2)}
- \left[ \Delta_1^{\scsc\#}\left(\frac{\partial L(..)}{\partial
y_1} \right)_{|..}\right]_{|(N_1^1+j,N_1^2)} \right. } \\[0.7cm]
{\ds \left. -\sqrt{\frac{N_1^2+1}{2}}\; \frac{\partial L(..)}{\partial
y_2}_{|(N_1^1+j, N_1^2+1)}\right\} \cdot h(N_1^1 +j, N_1^2) }
\end{array}
$$

$$
\begin{array}{l}
{\ds +\sum\limits_{j=1}^{N_2^2-N_1^1-1}
\left\{\frac{\partial L(..)}{\partial y}_{|(N_2^1, N_1^2+j)}
+ \sqrt{\frac{N_2^1}{2}}\; \frac{\partial L(..)}{\partial
y_1}_{|(N_2^1-1,N_1^2+j)} \right. } \\[0.6cm]
{\ds \left. - \left[\Delta_2^{\scsc\#}\left(\frac{\partial
L(..)}{\partial y_2}\right)_{|..}\right]_{|(N_2^1,N_1^2+j)}
\right\} \cdot h(N_2^1, N_1^2 + j) } \\[0.7cm]
{\ds +\sum\limits_{j=1}^{N_2^1-N_1^1-1}
\left\{\frac{\partial L(..)}{\partial y}_{|(N_1^1+j, N_2^2)}
- \left[\Delta_1^{\scsc\#}\left(\frac{\partial L(..)}{\partial
y_1}_{|..}\right)\right]_{|(N_1^1+j,N_2^2)} \right. } \\[0.7cm]
{\ds \left.+\sqrt{\frac{N_2^2}{2}}\; \frac{\partial L(..)}{\partial
y_2}_{|(N_1^1+j, N_2^2)}\right\} \cdot
h(N_1^1 +j, N_2^2) } \\[0.7cm]
{\ds +\sum\limits_{j=1}^{N_2^2-N_1^2-1}
\left\{\frac{\partial L(..)}{\partial y}_{|(N_1^1, N_1^2+j)}
-\sqrt{\frac{N_1^1+1}{2}}\; \frac{\partial L(..)}{\partial
y_1}_{|(N_1^1+1, N_1^2+j)} \right. } \\[0.7cm]
{\ds \left.- \left[\Delta_2^{\scsc\#}\left(\frac{\partial
L(..)}{\partial y_2}\right)_{|..}\right]_{|(N_1^1,N_1^2+j)}
\right\} \cdot h(N_1^1, N_1^2 + j) } \\[0.7cm]
{\ds + \left\{ \frac{\partial L(..)}{\partial y}_{|(N_1^1,N_1^2)}
- \frac{1}{\sqrt{2}} \left[\sqrt{N_1^1+1}\; \frac{\partial
L(..)}{\partial y_1}_{|(N_1^1+1,N_1^2)} \right.\right.}\\[0.7cm]
{\ds\left.\left.+ \sqrt{N_1^2+1}\; {\frac{\partial L(..)}{\partial
y_2}}_{|(N_1^1,N_1^2+1)}\right]\right\}\cdot h(N_1^1, N_1^2)}\\[0.7cm]
{\ds + \left\{ \frac{\partial L(..)}{\partial y}_{|(N_2^1,N_1^2)}
- \frac{1}{\sqrt{2}} \left[\sqrt{N_2^1} \frac{\partial
L(..)}{\partial y_{1}}_{|(N_2^1-1,N_1^2)} \right.\right.} \\[0.7cm]
{\ds \left.\left.- \sqrt{N_1^2+1}\; \frac{\partial L(..)}{\partial
y_{2}}_{|(N_2^1,N_1^2+1)}\right]\right\}\cdot h(N_2^1,N_1^2)}\\[0.7cm]
{\ds + \left\{ \frac{\partial L(..)}{\partial y}_{|(N_2^1,N_2^2)}
+ \frac{1}{\sqrt{2}} \left[\sqrt{N_2^1}\; \frac{\partial
L(..)}{\partial y_1}_{|(N_2^1-1,N_2^2)} \right.\right. }\\[0.7cm]
{\ds\left.\left.+ \sqrt{N_2^2}\; \frac{\partial L(..)}{\partial
y_2}_{|(N_2^1,N_2^2-1)}\right]\right\}\cdot h(N_2^1,N_2^2)}\\[0.7cm]
{\ds + \left\{ \frac{\partial L(..)}{\partial y}_{|(N_1^1,N_2^2)}
- \frac{1}{\sqrt{2}} \left[\sqrt{N_1^1+1}\; \frac{\partial
L(..)}{\partial y_{1}}_{|(N_1^1+1,N_2^2)} \right.\right.}\\[0.7cm]
{\ds\left.\left.- \sqrt{N_2^2}\; \frac{\partial L(..)}{\partial
y_{2}}_{|(N_1^1,N_2^2-1)}\right]\right\}\cdot h(N_1^1, N_2^2)\,.}
\end{array}
\eqno\raisebox{-57.5ex}{{\rm (A.I.7)}}
$$
\vskip0.6cm

The {\em imposed plus natural\/} boundary conditions [13]
imply that the boundary terms in (A.I.7) add up to zero. There
exist many possible boundary conditions. We shall adopt the
simplest of all, namely
$$
\begin{array}{rcl}
h(n^1, n^2)_{|{\mathrm{Boundary}}} &=& 0\,, \\[0.2cm]
\delta f(n^1, n^2)_{|{\mathrm{Boundary}}} &=& 0\,.
\end{array}
\eqno{\rm (A.I.8)}
$$

Note that the discrete boundary points in equation (A.I.7) are more
in number than those in equation (9). We have to prove now the
analogue of the Dubois-Reymond lemma [14].
\vskip2ex

\noindent
{\bf Lemma:} \ \ \,{\it Let the double sum
$$
\sum\limits_{n^1=N_1^1+1}^{N_2^1-1}\: \sum\limits_{n^2=N_1^2
+1}^{N_2^2-1}g (n^1,n^2) \,h(n^1, n^2) = 0
$$
for a function $g$ and an arbitrary function $h$ over the domain
$D\subset \N^2$ defined in equation {\rm (A.I.1)}. Then $g(n^1,n^2)
\equiv 0$ in $D.$ }
\vskip2ex

\noindent
{\bf Proof.} \ \ \,Choose $h(n^1,n^2) := \delta_{n^1}^{m^1}
\delta_{n^2}^{m^2}$ for some $(m^1,m^2) \in D.$ (The $\delta_n^m$
denotes the Kronecker delta.) Then
$$
\begin{array}{l}
{\ds \sum\limits_{n^1=N_1^1+1}^{N_2^1-1}\: \sum\limits_{n^2
=N_1^2+1}^{N_2^2-1} g (n^1,n^2) \,h(n^1, n^2) } \\[0.7cm]
{\ds \quad \; \, = \sum\limits_{n^1=N_1^1+1}^{N_2^1-1}\:
\sum\limits_{n^2=N_1^2+1}^{N_2^2-1}
g (n^1,n^2) \,\delta_{n^1}^{m^1} \delta_{n^2}^{m^2}
= g (m^1,m^2) = 0\vspace*{0.2cm}\,. }
\end{array}
$$
Since the choice of $(m^1,m^2) \in D$ is arbitrary, it follows that
$g(n^1,n^2) \equiv 0$ for all $(n^1,n^2) \in D.$
\hspace*{\fill}{\qed}
\vskip0.3cm

At this stage we have furnished essentially the proof of the following
generalization of the Euler-Lagrange theorem.
\vskip0.5cm

\noindent
{\bf Theorem (A.I.1):} \ \ \,{\it Let a function $f: \,D\subset \N^2
\rightarrow \R,$ where $D$ is defined in {\rm (A.I.1).} Let an action
functional ${\mathcal{A}}(f)$ be defined by {\rm (A.I.2).} The
stationary values of ${\mathcal{A}},$ under the boundary variation
$\delta f(n)_{|..} = 0,$ are given by the solutions of the partial
difference equation:
$$
\frac{\partial L(..)}{\partial y}_{|y=f(n), \,
y_A=\Delta_A^{\mbox{\tts\#}}f(n)} - \Delta_A^{\scsc\#} \left\{
\left[ \frac{\partial L(..)}{\partial y_A}\right]_{|y=f(n), \,
y_A=\Delta_A^{\mbox{\tts\#}}f(n)} \right\} = 0\vspace*{0.2cm}\,.
\eqno{\rm (A.I.9)}
$$
}
(Here, the index $A$ is summed over $\{1,2\}.$)
\vskip0.5cm

\section*{Appendix II: \, Partial difference and difference-differential 
conservation
equations}

Consider the discrete domain $D\subset \N^4$ given by
equation (A.I.1). Let $A^{\alpha ..} (n)$ be a real-valued
$(r\!+\!s)$-th order tensor field over $D\subset \N^4.$ (See
equations (34A,B).) The Lagrangian $L(y^{\alpha ..}\,;
y_\nu^{\alpha ..})$ is a real-valued, analytic, and relativistic
invariant function over a domain of the continuum $R^{4^{r+s}}\times
R^{4^{r+s+1}}.$ The action functional ${\mathcal{A}}(A^{\alpha ..})$
is given by the four-fold sum (compare equation (A.I.2))
$$
\begin{array}{rcl}
{\mathcal{A}} (A^{\alpha ..}) &:=& {\ds \sum\limits_{n^1
=N_1^1}^{N_2^1} \sum\limits_{n^2=N_1^2}^{N_2^2} \sum\limits_{n^3
=N_1^3}^{N_2^3} \sum\limits_{n^4=N_1^4}^{N_2^4} } \\[0.7cm]
&& L(y^{\alpha ..}\,; y_\nu^{\alpha ..})_{|y^{\alpha ..}
=A^{\alpha ..}(n), \, y_\nu^{\alpha ..}=
\Delta_\nu^{\mbox{\tts\#}}A^{\alpha ..}}\vspace*{0.2cm}\:.
\end{array}
\eqno{\rm (A.II.1)}
$$
In the usual case, $L$ is a polynomial function of the $4^{2(r+s)+1}$
real variables.

The Taylor expansion of $L$ is given by
$$
\begin{array}{l}
L(y^{\alpha ..} + \Delta y^{\alpha ..}, y_\nu^{\alpha ..}
+ \Delta y_\nu^{\alpha ..}) \\[0.4cm]
= L(y^{\alpha ..}, y_\nu^{\alpha ..})
+ {\ds \left[\frac{\partial L(..)}{\partial
y^{\alpha ..}}\Delta y^{\alpha ..} + \frac{\partial L(..)}{\partial
y_\nu^{\alpha ..}}\Delta y_\nu^{\alpha ..}\right] } \\[0.6cm]
+ {\ds \frac{1}{2} \left[\frac{\partial^2 L(..)}{\partial
y^{\alpha ..} \partial y^{\beta ..}} \:(\Delta y^{\alpha ..}) (\Delta
y^{\beta ..}) + \frac{\partial^2 L(..)}{\partial y^{\alpha ..}
\partial y_\nu^{\beta..}} \:(\Delta y^{\alpha ..}) (\Delta
y_\nu^{\beta..}) \right. } \\[0.6cm]
+ {\ds\left. \frac{\partial^2 L(..)}{\partial
y_\nu^{\alpha ..}\partial y_\sigma^{\beta..}} \:(\Delta
y_\nu^{\alpha ..}) (\Delta y_\sigma^{\beta..})\right]+ \ldots\;.}
\end{array} \vspace*{0.4cm}\,
\eqno{\rm (A.II.2)}
$$
Here we have followed the summation convention on every repeated Greek
index. Moreover, for a {\em second-degree\/} polynomial function $L,$
{\em the additional terms denoted by $\ldots$ exactly vanish.}

Now, we investigate the partial difference operations on $L(..).$
Using equations (4iii), (A.II.2), (5iv), (6i,ii,iii,iv) and
{\em not\/} summing the index $\mu ,$ we derive \vspace*{0.4cm}that
$$
\begin{array}{l}
\sqrt{2} \,\Delta_\mu^{\scsc\#} \Big[L(y^{\alpha ..}\,;y_\nu^{\alpha
..})_{|y^{\alpha ..}=A^{\alpha ..}(n), \, y_\nu^{\alpha
..}=\Delta_\nu^{\mbox{\tts\#}}A^{\alpha ..}}\Big] \\[0.3cm]
= (\sqrt{n^\mu\!+\!1}\,) \cdot L\Big\{A^{\alpha ..}
(..,n^\mu\!+\!1,..);\:2^{-1/2}\!\Big[\sqrt{n^\nu\!+\!1}\,
A^{\alpha ..}(..,n^\mu\!+\!1,..,n^\nu\!+\!1,..) \\[0.3cm]
\qquad - \sqrt{n^\nu} A^{\alpha ..} (..,n^\mu +1,..,n^\nu -1,..)
\Big]\Big\} \\[0.3cm]
- (\sqrt{n^\mu}\,) \cdot L\Big\{A^{\alpha ..}(..,n^\mu -1,..);
\:2^{-1/2}\Big[\sqrt{n^\nu+1}\,A^{\alpha ..}(..,n^\mu-1,..,n^\nu
+1,..) \\[0.3cm]
\qquad - \sqrt{n^\nu} A^{\alpha ..} (..,n^\mu -1,..,n^\nu
-1,..)\Big]\Big\}
\end{array}
$$
$$
\begin{array}{l}
= (\sqrt{n^\mu+1}\,) \cdot L\Big[A^{\alpha ..}(n) + \Delta_\mu
A^{\alpha ..}\,;\:\Delta_\nu^{\scsc\#} A^{\alpha ..}(n)+\Delta_\mu
\Delta_\nu^{\scsc\#} A^{\alpha ..}\Big] \\[0.3cm]
\qquad - (\sqrt{n^\mu}\,) \cdot L\Big[A^{\alpha ..}(n) -
\Delta_\mu^{\prime} A^{\alpha ..}\,;\:\Delta_\nu^{\scsc\#}
A^{\alpha ..}(n) - \Delta_\mu^{\prime}
\Delta_\nu^{\scsc\#} A^{\alpha ..}\Big] \\[0.3cm]
{\ds = \sqrt{n^\mu\!+\!1}\,\Biggr\{\!L(..)\!+
\!\!\left[\frac{\partial
L(..)}{\partial y^{\alpha ..}}_{|..}\!\cdot\!\Delta_\mu
A^{\alpha ..}\!+\!\frac{1}{2}\,\frac{\partial^2 L(..)}{\partial
y^{\alpha ..}\partial y^{\beta..}}_{|..}\!\!\cdot\!(\Delta_\mu
A^{\alpha ..}) (\Delta_\mu A^{\beta..})\!+\!..\right] }
\\[0.6cm]
{\ds + \Biggr[ \frac{\partial L(..)}{\partial
y_{\nu}^{\alpha ..}}_{|..} \cdot \Delta_\mu \Delta_\nu^{\scsc\#}
A^{\alpha ..} + \frac{1}{2}\,\frac{\partial^2 L(..)}{\partial
y_\nu^{\alpha ..} \partial y_{\sigma}^{\beta ..}}_{|..}\cdot
(\Delta_\mu \Delta_\nu^{\scsc\#} A^{\alpha ..})\,(\Delta_\mu
\Delta_\sigma^{\scsc\#} A^{\beta..})\Biggr] + ..\Biggr\} }\\[0.6cm]
{\ds - \sqrt{n^\mu}\,\Biggr\{\!L(..)+\Biggr[\!-\frac{\partial
L(..)}{\partial y^{\alpha ..}}_{|..}\!\cdot\!\Delta_\mu^{\prime}
A^{\alpha ..}\!+\!\frac{1}{2}\,\frac{\partial^2 L(..)}{\partial
y^{\alpha ..}\partial y^{\beta..}}_{|..}\!\cdot\!(\Delta_\mu^{\prime}
A^{\alpha..})(\Delta_\mu^{\prime} A^{\beta..})-..\Biggr] } \\[0.6cm]
{\ds + \Biggr[\!-\frac{\partial L(..)}{\partial
y_{\nu}^{\alpha ..}}_{|..}\!\cdot\!\Delta_\mu^{\prime}
\Delta_\sigma^{\scsc\#} A^{\alpha ..}\!+\!\frac{1}{2}\,
\frac{\partial^2 L(..)}{\partial y_\nu^{\alpha ..}
\partial y_{\sigma}^{\beta ..}}_{|..}\!\cdot\!(\Delta_\mu^{\prime}
\Delta_\nu^{\scsc\#} A^{\alpha ..})(\Delta_\mu^{\prime}
\Delta_\sigma^{\scsc\#} A^{\beta..})-..\Biggr]\Biggr\} }\\[0.6cm]
{\ds = \sqrt{2} \,[L(..)] \,\Delta_\mu^{\scsc\#}(1) + \frac{\partial
L(..)}{\partial y^{\alpha ..}}_{|..}\cdot \Big[\sqrt{n^\mu+1}
\,(\Delta_\mu A^{\alpha ..}) + \sqrt{n^\mu} \,(\Delta_\mu^{\prime}
A^{\alpha ..})\Big] } \\[0.5cm]
{\ds + \frac{\partial L(..)}{\partial y_{\nu}^{\alpha ..}}_{|..}
\cdot \Big[\sqrt{n^\mu +1} \,(\Delta_\mu \Delta_\nu^{\scsc\#}
A^{\alpha ..}) + \sqrt{n^\mu } \,(\Delta_\mu^{\prime}
\Delta_\nu^{\scsc\#} A^{\alpha ..})\Big] } \\[0.6cm]
{\ds + \frac{1}{2}\,\frac{\partial^2 L(..)}{\partial y^{\alpha ..}
\partial y^{\beta ..}}_{|..} \cdot \Big[\sqrt{n^\mu+1}
\,(\Delta_\mu A^{\alpha ..})(\Delta_\mu A^{\beta ..}) } \\[0.5cm]
\qquad - \sqrt{n^\mu } (\Delta_\mu^{\prime} A^{\alpha ..})
(\Delta_\mu^{\prime} A^{\beta ..})\Big]  \\[0.4cm]
{\ds + \frac{1}{2}\,\frac{\partial^2 L(..)}{(\partial
y_\nu^{\alpha..})\,(\partial y_{\sigma}^{\beta ..})}\cdot
\Big[\sqrt{n^\mu +1}\,(\Delta_\mu \Delta_\nu^{\scsc\#}
A^{\alpha ..})(\Delta_\mu \Delta_\sigma^{\scsc\#}
A^{\beta ..}) } \\[0.5cm]
\qquad - \sqrt{n^\mu } \,(\Delta_\mu^{\prime} \Delta_\nu^{\scsc\#}
A^{\alpha ..})\,(\Delta_\mu^{\prime} \Delta_\sigma^{\scsc\#}
A^{\beta ..})\Big] + .. \\[0.4cm]
{\ds = \sqrt{2} \,\Biggr\{[L(..)] \,\Delta_\mu^{\scsc\#}(1)
+ \frac{\partial L(..)}{\partial y^{\alpha ..}}_{|..}
\cdot [\Delta_\mu^{\scsc\#} A^{\alpha ..} - A^{\alpha ..}(n)
\Delta_\mu^{\scsc\#}(1)] } \\[0.6cm]
\qquad{\ds + \frac{\partial L(..)}{\partial
y_{\nu}^{\alpha ..}}_{|..} \cdot [\Delta_\mu^{\scsc\#}
\Delta_\nu^{\scsc\#} A^{\alpha ..} - (\Delta_\nu^{\scsc\#}
A^{\alpha ..}) \Delta_\mu^{\scsc\#}(1)] } \\[0.7cm]
{\ds + \frac{1}{2} \, \frac{\partial^2 L(..)}{\partial y^{\alpha..}
\partial y^{\beta ..}}_{|..}\cdot \big[\Delta_\mu^{\scsc\#}
(A^{\alpha ..}(n))\,(A^{\beta ..}(n)) - A^{\alpha ..}(n)
\Delta_\mu^{\scsc\#} A^{\beta ..} } \\[0.6cm]
\qquad - A^{\beta ..}(n) \Delta_\mu^{\scsc\#} A^{\alpha ..}
+ A^{\alpha ..}(n) A^{\beta ..}(n) \Delta_\mu^{\scsc\#}(1)\big]
\\[0.4cm]
{\ds + \frac{1}{2}\,\frac{\partial^2 L(..)}{\partial y_\nu^{\alpha..}
\partial y_{\sigma}^{\beta ..}}_{|..}\!\cdot\!\big[\Delta_\mu^{\scsc\#}
(\Delta_\nu^{\scsc\#} A^{\alpha ..}\!\cdot \Delta_\sigma^{\scsc\#}
A^{\beta ..})\!-\!(\Delta_\nu^{\scsc\#} A^{\alpha ..})
(\Delta_\mu^{\scsc\#}\Delta_\sigma^{\scsc\#} A^{\beta ..}) }\\[0.5cm]
- (\Delta_\sigma^{\scsc\#} A^{\beta ..})\,(\Delta_\mu^{\scsc\#}
\Delta_\nu^{\scsc\#} A^{\alpha..}) + (\Delta_\nu^{\scsc\#}
A^{\alpha..})\,(\Delta_\sigma^{\scsc\#} A^{\beta ..})
\Delta_\mu^{\scsc\#}(1)\big] + .. \Biggr\}.
\end{array}
\hspace*{\fill}\raisebox{-57ex}{\rm (A.II.3)}
$$
\vskip0.5cm

\noindent
Dividing this equation by $\sqrt{2},$ bringing the left-hand side
term \vspace*{0.1cm}to the right-hand side, and equating $\frac{\partial
L(..)}{\partial y^{\alpha ..}}_{|..} = \Delta_\nu^{\scsc\#} \Big[
\frac{\partial L(..)}{\partial y_\nu^{\alpha ..}}\Big]_{|..}$ by the
Euler-Lagrange equation (36A), we finally \vspace*{0.3cm}obtain:
$$
\begin{array}{l}
{\ds \left\{\left[\Delta_\nu^{\scsc\#}\!\left(\frac{\partial
L(..)}{\partial y_\nu^{\alpha ..}}\right)_{\!|..}\right]\!\!\cdot\!
\Delta_\mu^{\scsc\#} A^{\alpha ..}\!+\!\frac{\partial L(..)}{\partial
y_\nu^{\alpha ..}}_{|..} \Delta_\mu^{\scsc\#} \Delta_\nu^{\scsc\#}
A^{\alpha ..}\!-\!\Delta_\mu^{\scsc\#} [L(..)_{|..}]\right\}
 } \\[0.7cm]
{\ds + \frac{1}{2}\, \frac{\partial^2 L(..)}{\partial y^{\alpha..}
\partial y^{\beta ..}}_{|..} \cdot \big[\Delta_\mu^{\scsc\#}
(A^{\alpha ..}(n) \cdot A^{\beta ..}(n)) } \\[0.7cm]
\qquad - A^{\alpha ..} (n) \Delta_\mu^{\scsc\#} A^{\beta ..}
- (\Delta_\mu^{\scsc\#} A^{\alpha ..})\,(A^{\beta ..}(n))\big]
\\[0.45cm]
{\ds + \frac{1}{2}\, \frac{\partial^2 L(..)}{\partial y_\nu^{\alpha..}
\partial y_\sigma^{\beta ..}}_{|..} \cdot \big[\Delta_\mu^{\scsc\#}
(\Delta_\nu^{\scsc\#} A^{\alpha ..} \cdot \Delta_\sigma^{\scsc\#}
A^{\beta ..}) } \\[0.75cm]
\qquad - (\Delta_\nu^{\scsc\#} A^{\alpha ..})\, (\Delta_\mu^{\scsc\#}
\Delta_\sigma^{\scsc\#} A^{\beta ..}) - (\Delta_\mu^{\scsc\#}
\Delta_\nu^{\scsc\#} A^{\alpha ..})\,(\Delta_\sigma^{\scsc\#}
A^{\beta ..})\big] \\[0.5cm]
{\ds + [\Delta_\mu^{\scsc\#}(1)] \Biggr[L(..)_{|..}-\frac{\partial
L(..)}{\partial y^{\alpha ..}}_{|..} \cdot A^{\alpha ..}(n)
- \frac{\partial L(..)}{\partial y_\nu^{\alpha ..}}_{|..}
\cdot \Delta_\nu^{\scsc\#}A^{\alpha ..} } \\[0.7cm]
\qquad{\ds + \frac{1}{2}\, \frac{\partial^2 L(..)}{\partial y^{\alpha
..} \partial y^{\beta ..}}_{|..} \cdot A^{\alpha ..}(n) A^{\beta
..}(n) } \\[0.6cm]
{\ds + \frac{1}{2}\, \frac{\partial^2 L(..)}{\partial y_\nu^{\alpha..}
\partial {y_\sigma^{\beta ..}}}_{|..} \cdot (\Delta_\nu^{\scsc\#}
A^{\alpha ..})\,(\Delta_\sigma^{\scsc\#} A^{\beta ..})\Biggr]
+ ... = 0\,. }
\end{array} \vspace*{0.5cm}\,
\eqno\raisebox{-25.5ex}{\rm (A.II.4A)}
$$

\noindent
Here, the indices $\alpha ,\beta ,\nu ,\sigma $ are summed. We claim
that the above equations involving the relativistic operator
$\Delta_\mu^{\scsc\#} = iP_\mu$ constitute {\em relativistic\/}
conservation equations in the finite difference form. The
corresponding relativisitic difference-differential conservation
equations \vspace*{0.4cm}are
$$
\begin{array}{l}
{\ds \Biggr\{\Biggr[\Delta_b^{\scsc\#}\!\left(\frac{\partial
L(..)}{\partial y_b^{\alpha ..}}\right)_{|..}\!+\!\partial_t
\left(\frac{\partial L(..)}{\partial y_4^{\alpha ..}}\right)_{|..}
\Biggr]\!\cdot \Delta_a^{\scsc\#} A^{\alpha ..} + \frac{\partial
L(..)}{\partial y_{b}^{\alpha ..}}_{|..}\!\cdot \Delta_a^{\scsc\#}
\Delta_b^{\scsc\#} A^{\alpha ..} } \\[0.7cm]
\qquad{\ds + \frac{\partial L(..)}{\partial y_{4}^{\alpha ..}}_{|..}
\,\Delta_a^{\scsc\#} \partial_t A^{\alpha ..}
- \Delta_a^{\scsc\#} [L(..)_{|..}\,]\Biggr\} } \\[0.7cm]
{\ds + \frac{1}{2}\, \frac{\partial^2 L(..)}{\partial y^{\alpha ..}
\partial y^{\beta ..}}_{|..} \cdot \big[ \Delta_a^{\scsc\#}
(A^{\alpha ..} \cdot A^{\beta ..}) } \\[0.7cm]
\qquad - A^{\alpha ..} ({\bfm{n}},t)
\Delta_a^{\scsc\#} A^{\beta ..} - A^{\beta ..} ({\bfm{n}},t)
\Delta_a^{\scsc\#} A^{\alpha ..}\big]
\end{array}
$$
$$
\begin{array}{l}
{\ds + \frac{1}{2}\, \frac{\partial^2 L(..)}{\partial y_b^{\alpha ..}
\partial y_c^{\beta ..}}_{|..} \cdot \big[ \Delta_a^{\scsc\#}
(\Delta_b^{\scsc\#} A^{\alpha ..} \cdot \Delta_c^{\scsc\#}
A^{\beta ..}) } \\[0.7cm]
\qquad - (\Delta_b^{\scsc\#} A^{\alpha ..})\,
(\Delta_a^{\scsc\#} \Delta_c^{\scsc\#} A^{\beta ..}) -
(\Delta_c^{\scsc\#} A^{\beta ..})\,(\Delta_a^{\scsc\#}
\Delta_b^{\scsc\#} A^{\alpha ..})\big] \\[0.5cm]
{\ds + \frac{\partial^2 L(..)}{\partial y_b^{\alpha ..}
\partial y_4^{\beta ..}}_{|..} \cdot \big[ \Delta_a^{\scsc\#}
(\Delta_b^{\scsc\#} A^{\alpha ..} \cdot \partial_t
A^{\beta ..}) } \\[0.75cm]
\qquad - (\Delta_b^{\scsc\#} A^{\alpha ..})\,
(\Delta_a^{\scsc\#} \partial_t A^{\beta ..}) -
(\partial_t A^{\beta ..})\,(\Delta_a^{\scsc\#}
\Delta_b^{\scsc\#} A^{\alpha ..})\big] \\[0.55cm]
{\ds + \frac{1}{2}\, \frac{\partial^2 L(..)}{\partial y_4^{\alpha ..}
\partial y_4^{\beta ..}}_{|..} \cdot \big[ \Delta_a^{\scsc\#}
(\partial_t A^{\alpha ..} \cdot \partial_t
A^{\beta ..}) } \\[0.7cm]
\qquad - (\partial_t A^{\alpha ..})\,
(\Delta_a^{\scsc\#} \partial_t A^{\alpha ..}) -
(\partial_t A^{\beta ..})\,(\Delta_a^{\scsc\#}
\partial_t A^{\alpha ..})\big] \\[0.45cm]
{\ds + \Big[\Delta_a^{\scsc\#}(1)]\,[L(..)_{|..} - \frac{\partial
L(..)}{\partial y^{\alpha ..}}_{|..} \cdot A^{\alpha ..}
({\bfm{n}},t) } \\[0.65cm]
\qquad{\ds - \frac{\partial L(..)}{\partial y_b^{\alpha ..}}_{|..}
\cdot \Delta_b^{\scsc\#} A^{\alpha ..} - \frac{\partial
L(..)}{\partial y_4^{\alpha ..}}_{|..} \cdot \partial_t A^{\alpha
..} } \\[0.75cm]
{\ds + \frac{1}{2}\, \frac{\partial^2 L(..)}{\partial y^{\alpha ..}
\partial y^{\beta ..}}_{|..} \cdot A^{\alpha ..} ({\bfm{n}},t)
A^{\beta ..} ({\bfm{n}},t) } \\[0.75cm]
\qquad{\ds + \frac{1}{2}\, \frac{\partial^2 L(..)}{\partial
y_b^{\alpha..} \partial y_c^{\beta ..}}_{|..}\cdot (\Delta_b^{\scsc\#}
A^{\alpha ..})\,(\Delta_c^{\scsc\#} A^{\beta ..}) } \\[0.65cm]
{\ds + \frac{\partial^2 L(..)}{\partial y_b^{\alpha ..}
\partial y_4^{\beta ..}}_{|..} \cdot (\Delta_b^{\scsc\#}
A^{\alpha ..})\,(\partial_t A^{\beta ..}) } \\[0.7cm]
\qquad{\ds + \frac{1}{2}\, \frac{\partial^2 L(..)}{\partial
y_4^{\alpha ..} \partial y_4^{\beta ..}}_{|..}\cdot (\partial_t
A^{\alpha ..})\,(\partial_t A^{\beta ..})\Big] + \ldots = 0\,, }
\end{array}
\eqno\raisebox{-40.25ex}{\rm (A.II.4Bi)}
$$
\vskip4ex
$$
\!\begin{array}{l}
{\ds \Biggr\{\Biggr[\Delta_b^{\scsc\#}\!\Biggr(\frac{\partial
L(..)}{\partial y_b^{\alpha ..}}\Biggr)_{\!|..}\!+\partial_t\!
\Biggr(\frac{\partial L(..)}{\partial y_4^{\alpha ..}}\Biggr)_{\!|..}
\!\!\cdot\partial_t A^{\alpha ..}\!+\!\frac{\partial L(..)}{\partial
y_b^{\alpha ..}}_{|..}\Biggr]\!\!\cdot\partial_t \Delta_b^{\scsc\#}
A^{\alpha ..} } \\[0.7cm]
\qquad{\ds + \frac{\partial L(..)}{\partial y_4^{\alpha ..}}_{|..}
\cdot (\partial_t)^2 A^{\alpha ..} - \partial_t [L(..)_{|..}]\Biggr\}
+ \ldots = 0\,. }
\end{array} \vspace*{0.2cm}\,
\eqno{\rm (A.II.4Bii)}
$$
The above equations containing operators $\Delta_j^{\scsc\#} = iP_j$
and $\partial_t = iP_4$ on the same footing are {\em relativistic}. It
is extremely hard to put (A.II.4A) and (A.II.4Bi,ii) into the
relativistic conservation equations:
$$
\begin{array}{c}
\Delta_\nu^{\scsc\#} T_\mu^\nu = 0\,, \\[0.3cm]
\Delta_b^{\scsc\#} T_a^b + \partial_t T_a^4 = 0\,,
\quad \Delta_b^{\scsc\#} T_4^b + \partial_t T_4^4 = 0\vspace*{0.2cm}\,.
\end{array}
$$
However, using the equation (5v), the relativistic conservation
equation\linebreak (A.II.4A) can be cast into the \vspace*{0.2cm}form:
$$
\begin{array}{l}
{\ds (1 / \sqrt{2}) \Biggr\{\Biggr[\Delta_\nu \sqrt{n^\nu}
\frac{\partial L(..)}{\partial {y^{\alpha
..}_{\nu}}_{|(..,n^\nu-1,..)}}
\cdot \Delta_\mu^{\scsc\#} A^{\alpha ..} } \\[0.6cm]
{\ds + \frac{\partial L(..)}{\partial {y^{\alpha ..}_{\nu}}_{|..}}
\cdot (\Delta_\mu^{\scsc\#} A^{\alpha ..})_{|(..,n^{\nu}-1,..)} }
\\[0.6cm]
- \delta_\mu^\nu  L(..)_{|..}\Biggr]\Biggr\}
+ .. =: \Delta_\nu [T_\mu^\nu(n)] + .. = 0\,.
\end{array} \vspace*{0.3cm}\,
\eqno{\rm (A.II.5A)}
$$
Here, neither $T_\mu^\nu (n)$ are relativistic tensor components,
nor $\Delta_\mu$ is a relativistic difference-operator. {\em However,
the combination $\Delta_\nu T_\mu^\nu + ..$ \, a r e \, components
of a relativistic covariant vector!}

Similarly, from the relativistic difference-differential conservation
equations (A.II.4i,ii) we \vspace*{0.1cm}derive
$$
\begin{array}{l}
{\ds (1 / \sqrt{2}) \Delta_b \Biggr\{ \sqrt{n^b}\Biggr[\frac{\partial
L(..)}{\partial {y^{\alpha ..}_{b}}_{|(..,n^b-1,..)}} } \\[0.55cm]
{\ds + \frac{\partial L(..)}{\partial y_{b|..}^{\alpha ..}}
\cdot (\Delta_a^{\scsc\#} A^{\alpha ..})_{|(..,n^b-1,..)}
- \delta_b^a  L(..)_{|..}\Biggr]\Biggr\} } \\[0.7cm]
{\ds + \partial_t \Biggr\{\frac{\partial L(..)}{\partial {y_4^{\alpha
..}}_{|..}} \cdot \Delta_a^{\scsc\#} A^{\alpha ..}\Biggr\} + .. =:
\Delta_b [T_a^b] + \partial_t [T_a^4] + .. = 0\,, }
\end{array} \vspace*{0.5cm}\,
\eqno{\rm (A.II.5Bi)}
$$
\vskip0.5ex
$$
\begin{array}{l}
{\ds (1 / \sqrt{2}) \Delta_b \Biggr\{\sqrt{n^b}\Biggr[\frac{\partial
L(..)}{\partial {y^{\alpha ..}_{b}}_{|(..,n^b-1,..)}}\cdot\partial_t
A^{\alpha ..} } \\[0.6cm]
{\ds + \frac{\partial L(..)}{\partial {y_{b}^{\alpha ..}}_{|..}}
\cdot (\partial_t A^{\alpha ..})_{|(..,n^b-1,..)}\Biggr]\Biggr\}
 } \\[0.7cm]
{\ds + \partial_t \Biggr\{\frac{\partial L(..)}{\partial
y_{4|..}^{\alpha..}}\cdot \partial_t A^{\alpha ..}
- L(..)_{|..}\Biggr\} =:
\Delta_b T_4^b + \partial_t T_4^4 = 0\,.}
\end{array} \vspace*{0.4cm}\,
\eqno\raisebox{1ex}{\rm (A.II.5Bii)}
$$

We can generalize conservation equations (A.II.5A) and (A.II.5Bi,ii)
for a complex-valued tensor or spinor field $\phi^{..}(n)$ and
$\phi^{..} ({\bfm{n}},t)$ to the following \vspace*{0.2cm}equations:
$$
\begin{array}{l}
{\ds (1 / \sqrt{2}) \Delta_\nu \Biggr\{ \sqrt{n^\nu} \Biggr[
\frac{\partial L(..)}{\partial \rho_\nu^{..}}_{|(..,n^\nu-1,..)}
\cdot \Delta_\mu^{\scsc\#} \phi^{..} } \\[0.55cm]
{\ds + \frac{\partial L(..)}{\partial \rho_{\nu|..}^{..}}
\cdot (\Delta_\mu^{\scsc\#} \phi^{..})_{|(..,n^\nu-1,..)} + {\rm
(c.c.)} -\delta_\mu^\nu [L(..)]_{|..}\Biggr]\Biggr\} } \\[0.7cm]
{\ds - \sqrt{\frac{n^\mu}{2}} \Delta_\mu^{\prime} [L(..)]_{|..}
+ .. =: \Delta_\nu T_\mu^\nu + .. = 0\,,}
\end{array}
\eqno{\rm (A.II.6A)}
$$
\vskip3ex
$$
\begin{array}{l}
{\ds (1 / \sqrt{2}) \Delta_b \Biggr\{ n^b \Biggr[\frac{\partial
L(..)}{\partial \rho_{b|(..,n^b-1,..)}^{..}} \cdot
\Delta_a^{\scsc\#} \phi^{..}
+ \frac{\partial L(..)}{\partial \rho_{b}^{..}}_{|..} \,\cdot
} \\[0.7cm]
{\ds \cdot\, (\Delta_a^{\scsc\#} \phi^{..})_{|(..,n^b-1,..)} +
{\rm (c.c.)} -\delta_a^b [L(..)]_{|..}\Biggr]\Biggr\} }
{\ds - \sqrt{\frac{n^a}{2}} \Delta_a^{\prime} [L(..)]_{|..}
} \\[0.7cm]
{\ds + \partial_t \Biggr\{\Biggr[ \frac{\partial L(..)}{\partial
\rho_{4}^{..}}_{|..} \cdot \Delta_a^{\scsc\#} \phi^{..}
+ \frac{\partial L(..)}{\partial \overline{\rho_{4|..}^{..}}}
\cdot \Delta_a^{\scsc\#} \overline{\phi^{..}}\Biggr]\Biggr\} }
+ .. \\[0.75cm]
= \Delta_b T_a^b + \partial_t T_a^4 + .. = 0\,,
\end{array}
\eqno{\rm (A.II.6Bi)}
$$
\vskip4ex
$$
\begin{array}{l}
{\ds (1 / \sqrt{2}) \Delta_b \Biggr\{ \sqrt{n^b} \Biggr[
\frac{\partial L(..)}{\partial \rho_{b |(..,n^b-1,..)}^{..}}
\cdot \partial_t \phi^{..} } \\[0.6cm]
{\ds + \frac{\partial L(..)}{\partial \rho_{b |..}^{..}}
\cdot \partial_t \phi^{..}_{|(..,n^b-1,..)} + {\rm
(c.c.)} \Biggr]\Biggr\} } \\[0.7cm]
{\ds + \partial_t \Biggr\{ \frac{\partial L(..)}{\partial
\rho_{4|..}^{..}} \cdot \partial_t \phi^{..} +
\frac{\partial L(..)}{\partial \overline{\rho}_{4|..}^{..}} \cdot
\partial_t \overline{\phi^{..}} - [L(..)_{|..}]\Biggr\} } \\[0.75cm]
=: \Delta_b T_4^b + \partial_t T_4^4 = 0\,.
\end{array}
\eqno{\rm (A.II.6Bii)}
$$
\vskip3ex

\noindent
Here, (c.c.) indicates the complex-conjugation of the
{\em preceding\/} terms.

Now, we shall investigate the gauge invariance of the Lagrangian for
a complex-valued tensor or spinor field $\phi^{\alpha ..}(n)$ and
$\phi^{\alpha ..}({\bfm{n}},t)$ and the corresponding difference and
difference-differential conservation equations. A global,
infinitesimal gauge transformation is characterized by:

$$
\widehat{\phi}^{\alpha ..}(n) = [\exp (i\varepsilon)] \phi^{\alpha ..}
(n) = \phi^{\alpha ..}(n) + (i\varepsilon ) \phi^{\alpha..} (n) + 0
(\varepsilon^2)\vspace*{0.15cm},
\eqno{\rm (A.II.7A)}
$$
$$
\widehat{\phi}^{\alpha ..}({\bfm{n}},t) = [\exp (i\varepsilon)]
\phi^{\alpha ..}({\bfm{n}},t)
= \phi^{\alpha ..}({\bfm{n}},t) + (i\varepsilon) \phi^{\alpha ..}
({\bfm{n}},t) + 0(\varepsilon^2)\vspace*{0.4cm}.
\eqno{\rm (A.II.7B)}
$$
The invariance of the Lagrangian function under (A.II.7A)
\vspace*{0.2cm}implies
that
$$
\hspace*{-0.1cm}\begin{array}{rcl}
0 &=& L(\rho^{..}, \overline{\rho}^{..}\,; \rho_\mu^{..},
\overline{\rho}_\mu^{..})_{|\rho^{..} = \widehat{\phi}^{..}(n),
\,\overline{\rho}^{..} =
\overline{\widehat{\phi}}^{..}(n), \,\rho_\mu^{..} =
\Delta_\mu^{\mbox{\tts\#}} \widehat{\phi}^{..}, \,\overline{\rho}_\mu^{..}
= \Delta_\mu^{\mbox{\tts\#}} \overline{\widehat{\phi}}^{..}} \\[0.5cm]
&-& L(\rho^{..}, \overline{\rho}^{..}\,; \rho_\mu^{..},
\overline{\rho}_\mu^{..})_{|\rho^{..} = \phi^{..}(n),
\,\overline{\rho}^{..} =
\overline{\phi}^{..}(n), \,\rho_\mu^{..} =
\Delta_\mu^{\mbox{\tts\#}} \phi^{..}, \,\overline{\rho}_\mu^{..}
= \Delta_\mu^{\mbox{\tts\#}} \overline{\phi}^{..}} \\[0.5cm]
&=& {\ds (i \varepsilon) \Biggr[ \frac{\partial L(..)}{\partial
\rho^{..}} \rho^{..}\!-\!\frac{\partial L(..)}{\partial
\overline{\rho}^{..}} \overline{\rho}^{..}\!+\!\frac{\partial
L(..)}{\partial \rho_\mu^{..}} \rho_\mu ^{..}\!-\!\frac{\partial
L(..)}{\partial \overline{\rho}_\mu^{..}} \overline{\rho}_\mu^{..}
\Biggr]_{|..}\!+\!0 (\varepsilon^2). }
\end{array}
\eqno{\rm (A.II.8A)}
$$
\vskip0.8ex

\noindent
Dividing by $\varepsilon > 0,$ taking the limit $\varepsilon
\rightarrow 0_+$, and equating  $\frac{\partial L(..)}{\partial
{\rho^{..}}}_{|..} =\linebreak \Delta_\mu^{\scsc\#} \Big[
\frac{\partial L(..)}{\partial \rho_\mu^{..}}\Big]_{|..},$
we obtain \vspace*{0.2cm}that
$$
0 = i\Biggr\{\Biggr[ \Delta_\mu^{\scsc\#} \Biggr(\frac{\partial
L(..)}{\partial \rho_\mu^{\alpha ..}}\Biggr)_{|..}\Biggr] \cdot
\phi^{\alpha ..}(n) + \frac{\partial L(..)}{\partial {\rho_\mu^{\alpha
..}}_{|..}} \cdot \Delta_\mu^{\scsc\#} \phi^{\alpha ..}\Biggr\}+{\rm
(c.c.)}\vspace*{0.2cm}\,.
\eqno{\rm (A.II.9A)}
$$
Since $i\Delta_\mu^{\scsc\#} = P_\mu$ represent the relativistic
four-momentum operators, the equation (A.II.9A) incorporates the
{\em relativistic\/} partial difference equation for the
charge-current conservation. The difference-differential version
of the {\em relativistic\/} charge-current conservation is
furnished by:
$$
\begin{array}{l}
{\ds i\Biggr\{\Biggr[ \Delta_b^{\scsc\#} \Biggr(\frac{\partial
L(..)}{\partial \rho_b^{\alpha ..}}\Biggr)\Biggr] \cdot
\phi^{\alpha ..}({\bfm{n}},t) + \Biggr[ \partial_t \frac{\partial
L(..)}{\partial \rho_4^{\alpha..}}\Biggr)_{|..}\cdot \phi^{\alpha ..}
({\bfm{n}},t) } \\[0.7cm]
{\ds + \frac{\partial L(..)}{\partial {\rho_b^{\alpha ..}}_{|..}} \cdot
\Delta_b^{\scsc\#} \phi^{\alpha ..} ({\bfm{n}},t) + \frac{\partial
L(..)}{\partial {\rho_4^{\alpha ..}}_{|..}} \cdot \partial_t
\phi^{\alpha ..} ({\bfm{n}},t)\Biggr\} + {\rm (c.c.)} = 0\,. }
\end{array}
\eqno{\rm (A.II.9B)}
$$

\newpage

\thispagestyle{empty}

\ \\[-2.5cm]

$$ \begin{minipage}[t]{15cm}
\beginpicture
\setcoordinatesystem units <1truecm,1truecm>
\setplotarea x from 0 to 13, y from 10 to -10

\setsolid
\setlinear
\plot 0 0 12 0 /
\plot 6 6 6 -6 /
\put {\vector(1,0){4}} [Bl] at 12 0
\put {\vector(0,1){4}} [Bl] at 6 6

\setquadratic
\circulararc 360 degrees from 5.35 0 center at 6 0
\circulararc 360 degrees from 4.75 0 center at 6 0
\circulararc 360 degrees from 4.15 0 center at 6 0

\setlinear
\plot 6 0  5.5 -0.3775 /
\plot 6 0  6.75 -1 /
\plot 6 0  7.075 1.5 /

\put {${\rm p}^\mu$} at 6 6.4
\put {${\rm q}^\mu$} at 12.5 0
\put {$\scriptstyle 1$} at 5.8725 -0.35
\put {$\scriptstyle \sqrt{3}$} at 6.75 -0.55
\put {$\scriptstyle \sqrt{5}$} at 7.05 1.075

\put {{\bf FIG. 1} \ \ One discrete phase plane.} at 6 -9

\endpicture
\end{minipage} $$

\newpage

\thispagestyle{empty}

\ \\

$$ \hspace*{-1cm}\begin{minipage}[t]{15cm}
\beginpicture
\setcoordinatesystem units <0.8truecm,0.8truecm>
\setplotarea x from 0 to 15, y from 10 to -10

\setsolid
\setlinear
\plot 0 0 14 0 /
\plot 6 6 6 -6 /
\put {\vector(1,0){4}} [Bl] at 14 0
\put {\vector(0,1){4}} [Bl] at 6 6

\setquadratic
\plot 7.85 1.1  7.8 0.8  8 0.5 /
\plot 8 0.5  8.15 0.2  8.15 -0.4 /
\plot 13.05 5.35 13.4 5.5 13.7 5.575 /

\setlinear
\put {\vector(1,0){4}} [Bl] at 13.7 5.575
\put {\vector(0,-1){4}} [Bl] at 8.15 -0.4

\put {${\rm n}^2$} at 6 6.5
\put {${\rm n}^1$} at 14.5 0.1
\put {$({\rm N}_1^1, {\rm N}_1^2)$} at 9.25 -0.9
\put {$({\rm N}_2^1, {\rm N}_2^2)$} at 14.8 5.575

\setplotsymbol ({\circle*{3.5}} [Bl])
\plotsymbolspacing13mm
\setdots <5mm> \plot 8 5.2  13.5 5.2 /
\plot 8 4.2  13.5 4.2 /
\plot 8 3.2  13.5 3.2 /
\plot 8 2.2  13.5 2.2 /
\plot 8 1.2  13.5 1.2 /

\setsolid

\put {{\bf FIG. 2} \ \ A two-dimensional dicrete domain.} at 8 -9

\endpicture
\end{minipage} $$

\end{document}